\documentclass[%
amsmath,aps,showkeys,
amssymb,twocolumn,pra,
superscriptaddress,
]{revtex4-2}

\usepackage[T1]{fontenc}

\usepackage{amsmath, amssymb, physics}
\usepackage{geometry}
\usepackage{braket}
\usepackage{graphicx}
\usepackage{subfigure}
\usepackage{booktabs}
\usepackage{color}
\usepackage[dvipsnames]{xcolor}
\usepackage[colorlinks=true,citecolor=NavyBlue,linkcolor=RubineRed,urlcolor=Cerulean,pdfencoding=auto]{hyperref}

\begin{document}

\title{Modelling the Impact of Device Imperfections on Electron Shuttling in SiMOS devices.}

\author{Jack J. Turner}
\affiliation{Quantum Motion, 9 Sterling Way, London N7 9HJ, United Kingdom}

\author{Christian W. Binder}
\affiliation{Department of Materials, University of Oxford, Parks Rd, Oxford OX1 3PJ, United Kingdom}
\affiliation{Quantum Motion, 9 Sterling Way, London N7 9HJ, United Kingdom}

\author{Guido Burkard}
\affiliation{Department of Physics, University of Konstanz, D-78457 Konstanz, Germany}

\author{Andrew J. Fisher}
\affiliation{Dept. of Physics and Astronomy and London Centre for Nanotechnology,
University College London, London WC1E 6BT, United Kingdom}
\affiliation{Quantum Motion, 9 Sterling Way, London N7 9HJ, United Kingdom}

\begin{abstract}
Extensive theoretical and experimental work has established high-fidelity electron shuttling in Si/SiGe systems, whereas demonstrations in Si/SiO$_2$, also known as silicon metal-oxide-semiconductor (SiMOS), devices remain at an early stage. To help address this, we perform full 3D simulations of conveyor-belt charge shuttling in a realistic SiMOS device, building on earlier 2D modelling. We solve the Poisson and time-dependent Schrödinger equations for varying shuttling speeds and gate voltages, focusing on potential pitfalls of typical SiMOS devices such as oxide–interface roughness, gate fabrication imperfections, and charge defects along the transport path. The simulations reveal that for low clavier-gate voltages, the additional oxide screening in multi-layer gate architectures causes conveyor-belt shuttling to collapse to the bucket-brigade mode, inducing considerable orbital excitation in the process. Increasing the confinement restores conveyor-belt operation, which we find to be robust against interface roughness, gate misalignment, and charge defects buried in the oxide. However, our results indicate that defects located at the Si/SiO$_2$-interface can induce significant orbital excitation. For lower conveyor gate biases, positive defects in the transport channel can even capture passing electrons. Hence we identify key challenges and find operating regimes for reliable charge transport in SiMOS architectures.

\end{abstract}

\maketitle

\section{Introduction}
\label{sec:intro}
Silicon-based spin qubits have emerged as a strong candidate for scalable quantum architectures due to their high packing density \cite{Zwerver2022qubitsmadebysemiconductor}, long coherence times \cite{tyryshkin2012electronspincoherence, zwaneburg2013siliconquantumelectronics}, and compatibility with industry fabrication processes \cite{Gonzalez-Zalba2021scalingsilicon-based, Vandersypen2017interfacingspinqubits}. High fidelity single-qubit  \cite{Yoneda2018quantumdotspinqubit, Veldhorst2014addressablequantumdot} and two-qubit gates \cite{Noiri2022fastuniversalquantum, Xue2022quantumlogicwithspin, Mills2022twoqubitsilicon} --- realised by electron spin resonance (ESR) or electric dipole spin resonance (EDSR) and the electron exchange interaction \cite{burkard2023semiconductorspinqubits} --- as well as initialisation \cite{mills2022high-fidelity,huang2024high} and fast, compact readout \cite{swift2025superinductordeepsubmicronintegrated, clarke2025spinreadout22nm} have been demonstrated.

\begin{figure*}[t]
    \centering
    \includegraphics[width=1\textwidth]{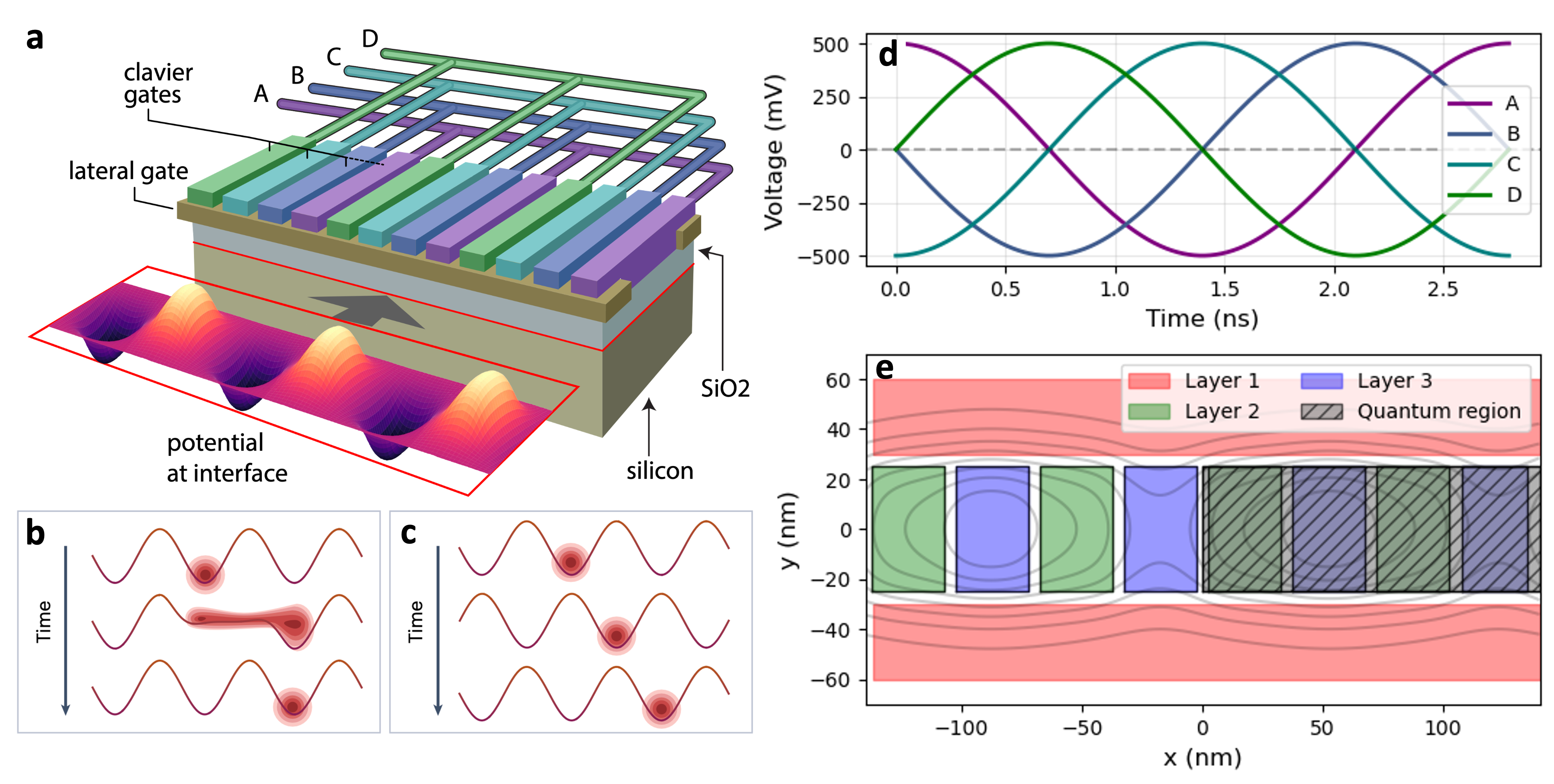}\label{fig: combined schematic}
    \caption{(a) Schematic of a typical quantum dot array structure used for shuttling with clavier-gates connected in ABCDABCD configuration. Figure modified from \cite{jeon2025robustness}. (b) Bucket-brigade shuttling transfers the electron wavefunction between adjacent QDs through a series of adiabatic Landau-Zener transitions that tip the particle from one dot to the next. (c) Conveyor-belt shuttling creates a smoothly travelling potential minimum by applying phase-shifted sinusoidal voltages to adjacent electrodes, continuously transporting the electron along the array. (d) Example phase-shifted voltage pulses applied to clavier-gates for $V_c=500$~mV and shuttling speed $v=50$~m/s, which corresponds to a frequency of 357~MHz for the 4-gate unit cell with period $d=140~\text{nm}$ shown in~(e). (e) Cross-section of the simulated shuttling device, with its clavier-gates in ABCDABCD configuration. The layer 1 confinement gates (`side-gates') are separated from the Si/SiO$_2$ interface in the z-direction by 15~nm of oxide, and each layer has 5~nm of oxide deposited between them. The electrostatic potential is solved for the entire device and plotted here with contours. The Schrödinger evolution occurs in a smaller quantum region in the channel, shaded in gray.}
    \label{fig:shuttling_schematic}
\end{figure*}

Owing to the signal fan-out problem, current state-of-the-art devices are mostly restricted to linear, bilinear, or trilinear layouts \cite{steinacker2024300mmfoundrysilicon, George202512spinqubitarrays, li2025trilinearquantumdotarchitecture}, and a central challenge is establishing longer-range connectivity between qubits whilst preserving coherence. Shuttling --- physically transporting electrons around a silicon chip using an architecture like that shown in 
Fig.~\ref{fig:shuttling_schematic}a --- is an empowering protocol for facilitating quantum information transfer across large distances, enabling enhanced or even all-to-all connectivity \cite{langrock2023blueprint}. This is valuable both for noisy intermediate-scale quantum (NISQ) applications and for longer-term fault-tolerant quantum computation (FTQC) approaches: for NISQ it can significantly reduce circuit depth by eliminating the need for SWAP operations required to bring distant qubits into proximity for two-qubit gates, while for FTQC, shuttling-based proposals supporting the surface code \cite{Kunne2024spinbus, taylor2005, cai2023loopedpipelines, siegel2025snakesplane} offer promising alternatives to SWAP-based architectures. By exploiting the long-range connectivity enabled by shuttling, these proposals have also been extended to support high-rate schemes such as low-density parity-check codes \cite{siegel2xn}. Such schemes rely primarily on conveyor-belt shuttling, which was shown to be superior to bucket-brigade shuttling in an analysis by Langrock \textit{et al.} \cite{langrock2023blueprint}. The bucket-brigade scheme (Figure \ref{fig:shuttling_schematic}b) uses a series of adiabatic Landau-Zener transitions (LZT) across an array of tunnel-coupled quantum dots (QDs) to transport electrons by tipping the electron from one well to the next \cite{ginzel2020, mills2019shuttling}, while the conveyor-belt protocol  (Fig.~\ref{fig:shuttling_schematic}c) transports electrons in a single moving quantum dot \cite{seidler2022conveyor}. Both methods require only a fixed number of voltage signals, but for bucket-brigade shuttling these must be carefully fine-tuned to trigger only adiabatic LZTs in specific QDs, since any diabatic LZT will cause the shuttling direction to reverse. For these reasons conveyor-belt shuttling is generally considered the preferred method and is what we consider in this work.

In silicon-germanium (Si/SiGe) heterostructures, recent experimental progress has demonstrated successful charge and spin shuttling \cite{seidler2022conveyor, mills2019shuttling, Yoneda2021coherentspinqubittransport, Xue2024QuBus}, most notably by De Smet \textit{et al.} \cite{desmet2025highfidelity} who used conveyor-belt mode to transport an electron over 10~\text{$\mu m$} in under 200~ns with 99.5$\%$ fidelity. Volmer \textit{et al.} \cite{volmer2026mappinggfactorscomplexintervalley} took this further, using electron shuttling to map out the g-factor and complex valley coupling within a 40~nm $\times$ 400~nm channel. These milestones prove that coherent transport is increasingly reliable, recently culminating in the use of shuttling to implement weight-four parity checks and generate five-qubit GHZ states \cite{undseth2026weightfourparitycheckssilicon}. Such achievements highlight the practical potential for shuttling-based quantum processors, paving the way for scalable FTQC.

However, translating these techniques to silicon oxide (Si/SiO$_2$)-based devices, also known as SiMOS devices, introduces additional complexities. Unlike in Si/SiGe-based devices, where a germanium-alloyed buffer layer of silicon separates trapped electrons from the oxide layer, SiMOS devices have electrons confined directly against the abrupt heterointerface. Without the buffer layer, confined electrons do not benefit from the same enhanced electrostatic screening of defects in the oxide layer and at the Si/SiO$_2$ interface. As a result, they may be expected to be more sensitive to irregularities in the electrostatic potential landscape caused by the amorphous oxide layer. Despite these difficulties, SiMOS devices offer key advantages: they are compatible with standard semiconductor fabrication processes \cite{Gonzalez-Zalba2021scalingsilicon-based}, have larger spatial variations in the g-factor that can enable electrostatic control without external micromagnets \cite{burkard2023semiconductorspinqubits}, and typically have larger and more tunable valley splittings due to the abrupt interface and larger conduction band offset \cite{yang2013spinvalleylifetimes, gamble2016valleysplittingofsingleelectron, Petit2018spinlifetimeandchargenoise}. Therefore, establishing the reliability of electron shuttling in Si/SiO$_2$-based devices is an important objective, and it is encouraging that successful charge shuttling experiments are beginning to emerge for modest distances. For instance, Lin \textit{et al.} \cite{lin2025interplayzeemansplittingtunnel} recently demonstrated high-fidelity bucket-brigade shuttling across three QD sites on a SiMOS device, utilising Pauli spin-blockade (PSB) readout to achieve an average spin shuttling fidelity of 99.8\%.

Charge defects buried in the oxide or located at the Si/SiO$_2$ interface are known to cause transistor threshold voltage shifts \cite{campbell2007atomicscaledefects}, cause random telegraph fluctuations ($1/f$ noise)  \cite{kirton1989noiseinsolidstatemicrostrucutres, kirton1989individualdefects, payne2015atomicresolutionsinglespin, fleetwood2023interfacetraps}, and increase qubit-to-qubit variability \cite{martinez2024mitigatingvariabilityepitaxialheterostructurebasedspinqubit}. They are also expected to present the greatest obstacle to successful charge shuttling in SiMOS devices. These defects have been studied with a wide range of techniques including transport measurements \cite{fleetwood2023interfacetraps, koh1997quanitativecharacterisationofsi-sio2interfacetraps}, electron spin resonance \cite{Brower1982defectsandimpurities, Lenahan2002radiationinducedinterfacetraps,Poindexter1984electrontrapsandpbcenters}, and frequency-modulated atomic force microscopy \cite{cowie2024spatiallyresolveddielectricloss, czarnecki2025hydrogenpassivationeffects}. Charge defects in the SiO$_2$ layer are most commonly identified as oxygen vacancies where neighbouring Si atoms form a direct bond, or as $E'$ centers associated with an unpaired electron on a silicon atom bonded to three (rather than four) oxygen atoms, although additional species have also been proposed \cite{wilhelmerabinitioinvestigationsinamorphoussilicondioxide, conleyelectronspinresonance, goes2018identificationofoxidedefects, helms1994thesiliconsilicondioxidesystem}. Interface defects, on the other hand, arise from the lattice mismatch between the silicon and oxide crystals, and at the atomic layer are understood to be silicon dangling bonds --- trivalent silicon atoms with an unpaired valence electron \cite{helms1994thesiliconsilicondioxidesystem}. These are identified as $P_{b0}$ and $P_{b1}$ centers, which possess energy levels that fall within the silicon band gap and are amphoteric in nature, acting as both electron donors or acceptors with the silicon bulk \cite{kato2006originofp_b1center, Lenahan2002directexperimentalevidence, ragnarsson2002electricalcharacterisationofpbcenters}. Understanding the impact of these traps on electrons during shuttling is therefore critical for the development of SiMOS quantum computing.

Fortunately, a previous 2D model by Jeon \textit{et al.} \cite{jeon2025robustness} showed that, even with as few as three independent gate electrodes, charge shuttling remained robust against nearby negatively charged defects, as well as against pulse imperfections and Johnson-Nyquist noise. Indeed, only a substantial density of defects near the center of the channel was able to prevent electron shuttling. Whilst encouraging, this study neglected the impact of positive charge traps, three-dimensional effects, and interface roughness, motivating the work presented here. By addressing these remaining questions, we aim to provide conclusive numerical evidence in support of successful conveyor-belt charge shuttling in SiMOS devices.

In order to obtain physically accurate results for charge shuttling in SiMOS, we solve the Poisson equation for a realistic shuttling device to determine the potential landscape at different times. Subsequently, we solve the time-dependent Schrödinger equation to track electron dynamics during shuttling. By simulating different shuttling scenarios, we assess the impact of gate voltages and shuttling speeds, interface roughness, gate imperfections, and both negative and positive charge traps. In particular, we identify a transition from the conveyor-belt mode of shuttling operation to bucket-brigade caused by the multi-layer gate structure. Moreover, we find that the existence of positive charge traps in the shuttling channel would pose a significant challenge, potentially leading to bound states that prevent successful shuttling.

This work focuses on the electron's orbital degrees of freedom, which establishes the prerequisite bounds on electron transport reliability in SiMOS devices. This crucial first step provides the foundation for understanding spin and valley dynamics during electron shuttling, which is ultimately necessary to assess the quality of quantum information transfer.

\section{Model}

\subsection{Shuttling Device}
\label{sec: shuttling device}

We perform simulations using an idealised model shuttling device. The device consists of a pair of side-gates that provide the lateral confinement of the shuttling channel and a set of clavier-gates arranged in a repeating four-gate unit cell. The side-gates are separated from the Si/SiO$_2$ interface in the z-direction by 15 nm of oxide, and each additional gate layer has 5 nm of oxide between them thereby reproducing realistic deposition processes \cite{Li2020aflexible300nm}.

We apply standard, phase-shifted voltages, like those shown in Fig.~\ref{fig:shuttling_schematic}d, without voltage offsets to the clavier-gates in order to generate the time-dependant potential for conveyor-belt shuttling. The side-gate voltages are held constant, while the voltage applied to the $k$-th clavier-gate ($k=0,1,2,3$) at time $t$ is given by
\begin{equation}\label{eq: V_clavier}
    V_{\mathrm{clav},k}(t)
    = V_{c}\,\sin\!\left(\omega t + \frac{\pi}{2}\, k\right),
\end{equation}
where the angular frequency is related to the shuttling speed and distance by $\omega=2\pi v/d$.

\subsection{Interface Roughness}
\label{sec: interface_roughness}

During device fabrication, the deposition process does not produce perfectly flat interfaces between crystal layers --- there is multi-scale roughness down to the atomic scale \cite{Cifuentes2024boundsto}. We characterise this roughness using the autocorrelation function (ACF) and its Fourier transform, the power spectral density (PSD). The ACF describes the spatial coherence of roughness at different separations, capturing whether nearby points on the interface tend to have similar heights $h(\mathbf{r})$ (positive correlation) or whether the height varies randomly with distance:
\begin{align}
    \text{ACF}(\mathbf{r}) &= \langle h(\mathbf{r}_0) h(\mathbf{r}_0 + \mathbf{r}) \rangle.
\end{align}
The PSD provides the reciprocal space representation of the roughness, decomposing the interface variations into their constituent spatial frequency (wavevector) components and quantifying which length scales dominate the roughness spectrum:
\begin{align}
    \text{PSD}(\mathbf{q}) &= \mathcal{F}\{\text{ACF}(\mathbf{r})\} = \int d^3r \text{ACF}(\mathbf{r}) e^{-i\mathbf{q}\cdot\mathbf{r}}.
\end{align}
In an older study, Goodnick \textit{et al.} \cite{goodnick1985surface}, performed transmission electron microscopy on Si/SiO$_2$ samples and found that exponential-form ACFs provided the best model. However, more recent data motivates a power-law functional form for the ACF that more accurately describes the self-affine structure of the roughness \cite{yoshinobu1995scaling}. Hence we take
\begin{align}
    \text{PSD}(\mathbf{q}) \propto q^{-2(1+H)},
\end{align}
where $H$ is the Hurst coefficient, which is related to the fractal dimension $D$ by $H=3-D$.

To generate individual realisations of the rough interface, random phases are assigned to each Fourier mode $\mathbf{q}$ with amplitudes determined by the PSD, and the spectrum is normalised to achieve a given root-mean-square (RMS) roughness. Finally, we apply the inverse Fourier transform to produce a field of surface heights with the desired statistics.

\subsection{Potential Landscape}
\label{sec: potential landscape}

The total device potential is decomposed as
\begin{equation}\label{eq: total potential}
    V(\mathbf{r}, t)
    = V_{g}(\mathbf{r}, t)
    + V_{\mathrm{ox}}(\mathbf{r})
    + V_{\mathrm{ch}}(\mathbf{r}),
\end{equation}
where $V_{\mathrm{ox}}(\mathbf{r})$ is the additional oxide potential, caused by the conduction band offset. We model this as
\begin{equation}
    V_{\mathrm{ox}}(\mathbf{r})
    = 3000~\mathrm{meV}\,\Theta(z - h(x,y)),
\end{equation}
where $h(x,y)$ is the local height of the Si/SiO$_2$ interface and $\Theta$ the Heaviside step function. The term $V_{\mathrm{ch}}(\mathbf{r})$ accounts for the electrostatic contribution from localised charge traps,  which we model as point charges each carrying one elementary charge $q=\pm e$.

$V_{g}(\mathbf{r}, t)$ is the gate-generated electrostatic potential, which is computed at each time step by solving the Poisson equation for the shuttling device with applied gate voltages given by Eq.~\eqref{eq: V_clavier}:
\begin{equation}
\nabla \cdot \left[ \epsilon(\mathbf{r}) \nabla \Phi(\mathbf{r}) \right] = -\rho(\mathbf{r}),
\end{equation}
where $\epsilon(\mathbf{r})$ is the spatially varying permittivity and $\rho(\mathbf{r})$ is the charge density. Because the equation is linear in both the boundary conditions and the source terms, the total potential landscape can be efficiently assembled from a set of precomputed Green’s functions. Further details on this approach are given in Appendix~\ref{Appendix: poisson}.

\subsection{Schr\"odinger Evolution}
\label{sec: schrodinger evolution}

To simulate electron dynamics according to the time-dependent Schr\"odinger equation
\begin{equation}
i\hbar \frac{\partial}{\partial t}\ket{\psi(t)} = H(t)\ket{\psi(t)},
\end{equation}
we use the single-valley effective-mass Hamiltonian
\begin{equation}
\label{eq: Hamiltonian}
    H(t) = -\frac{\hbar^{2}}{2}\,\nabla \cdot \mathbf{m}^{-1} \nabla + V(\mathbf{r}, t),
\end{equation}
where $\mathbf{m}^{-1}=\mathrm{diag}(m_{\ell}^{-1},m_{\ell}^{-1},m_t^{-1})$ is the inverse effective-mass tensor with silicon transverse and longitudinal masses $m_{t}=0.190\,m_{e}$ and $m_{\ell}=0.916\,m_{e}$, respectively. The time-dependent electrostatic potential $V(\mathbf{r},t)$ is assembled according to Eq.~\eqref{eq: total potential}.

The workhorse of our numerical simulations is the \emph{spectral projection method}, the implementation of which is presented in Appendix~\ref{Appendix: schrodinger} along with its derivation in Appendix~\ref{Appendix: subspace derivation}. This approach separates the rapidly oscillating dynamical phase from the slowly varying amplitudes associated with the system’s response to the evolving potential landscape. At each timestep, we compute the instantaneous eigenstates $\ket{\phi_n(t)}$ and eigenenergies $\epsilon_n(t)$ of $H(t)$ using a Lanczos eigensolver, and expand the state as $\ket{\psi(t)} = \sum_n c_n(t)\ket{\phi_n(t)}$. The evolution is then approximated by
\begin{align}
    c_n(t+dt)
    = \sum_m O_{nm}\, e^{-i\epsilon_m(t)\,dt/\hbar}\, c_m(t),
\end{align}
where 
\begin{equation}
O_{nm} = \braket{\phi_n(\mathbf{r},t+dt) | \phi_m(\mathbf{r},t)}
\end{equation}
is the overlap matrix between consecutive instantaneous eigenbases.  
This method allows for comparatively large timesteps while accurately capturing orbital excitations and level mixing during shuttling.

To validate the accuracy of the projection method, we benchmarked it against a direct split-operator Trotter evolution, which offers a high-accuracy but computationally expensive reference. The results, presented in Figure \ref{fig: neg defect validation} of Appendix \ref{Appendix: schrodinger}, demonstrate that the two methods agree closely across a wide range of shuttling conditions, providing confidence that the projection method accurately captures the relevant dynamics while enabling extensive parameter sweeps with manageable computational cost.

For all numerical experiments, the initial wavefunction $\ket{\psi(0)}$ is chosen as the instantaneous ground state at $t=0$, multiplied by a plane-wave factor $e^{-ipx/\hbar}$ to impart a starting velocity $v = p/m_t$.  This prevents sloshing in the confinement potential that would otherwise arise from beginning with a stationary electron.

For all the simulations presented in this study, we leveraged GPU-accelerated FFT and Lanzcos operations implemented through the CuPy Python library \cite{cupy_learningsys2017} and run on a Nvidia T4 Tensor.

\section{Results}

As discussed in Sec.~\ref{sec:intro}, we evaluate the impact of gate voltages and shuttling speeds, oxide-interface roughness, gate fabrication imperfections, and charge defects on the quality of electron shuttling. We define a complete shuttle as one full period of the quantum simulation domain ($d=140~\text{nm}$), and characterise the quality of the process using two key metrics: (1) the probability of electron loss from the conveyor $P_L$, and (2) the degree of adiabaticity, quantified by the fidelity of the shuttled wavefunction with the instantaneous ground state, $\mathcal{F}=|c_0(t)|^2$.

\subsection{Impact of Low Confinement}
\label{sec: results/low confinement}

\begin{figure*}[p]
    \centering
    \includegraphics[width=1\textwidth]{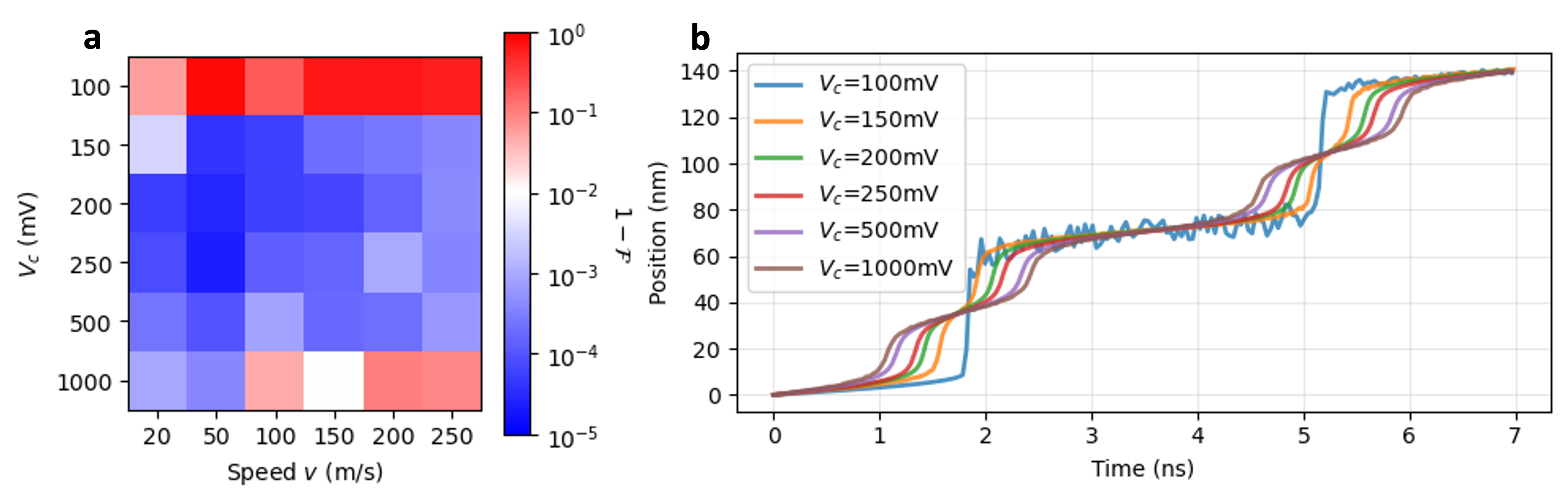}
    \caption{(a) The orbital ground state infidelity at the end of the shuttling down the channel for different clavier-gate voltages and shuttling speeds, with the side-gate voltage fixed at $V_s=-1500~\text{mV}$. A diverging colourmap centered on a ground state fidelity of 99\% ($1-\mathcal{F}=10^{-2}$) indicates where shuttling becomes adiabatic. (b) Mean position of electrons shuttled at 20~m/s along a flat interface for different clavier-gate voltages with $V_s=-1500~\text{mV}$. At $V_c=100~\text{mV}$ the confinement under layer 3 gates is very weak due to additional screening and the conveyor minimum effectively jumps from under gate C to under gate A, causing orbital excitations.}
    \label{fig:low_confinement}
\end{figure*}

\begin{figure*}[p]
    \centering
    \includegraphics[width=1\textwidth]{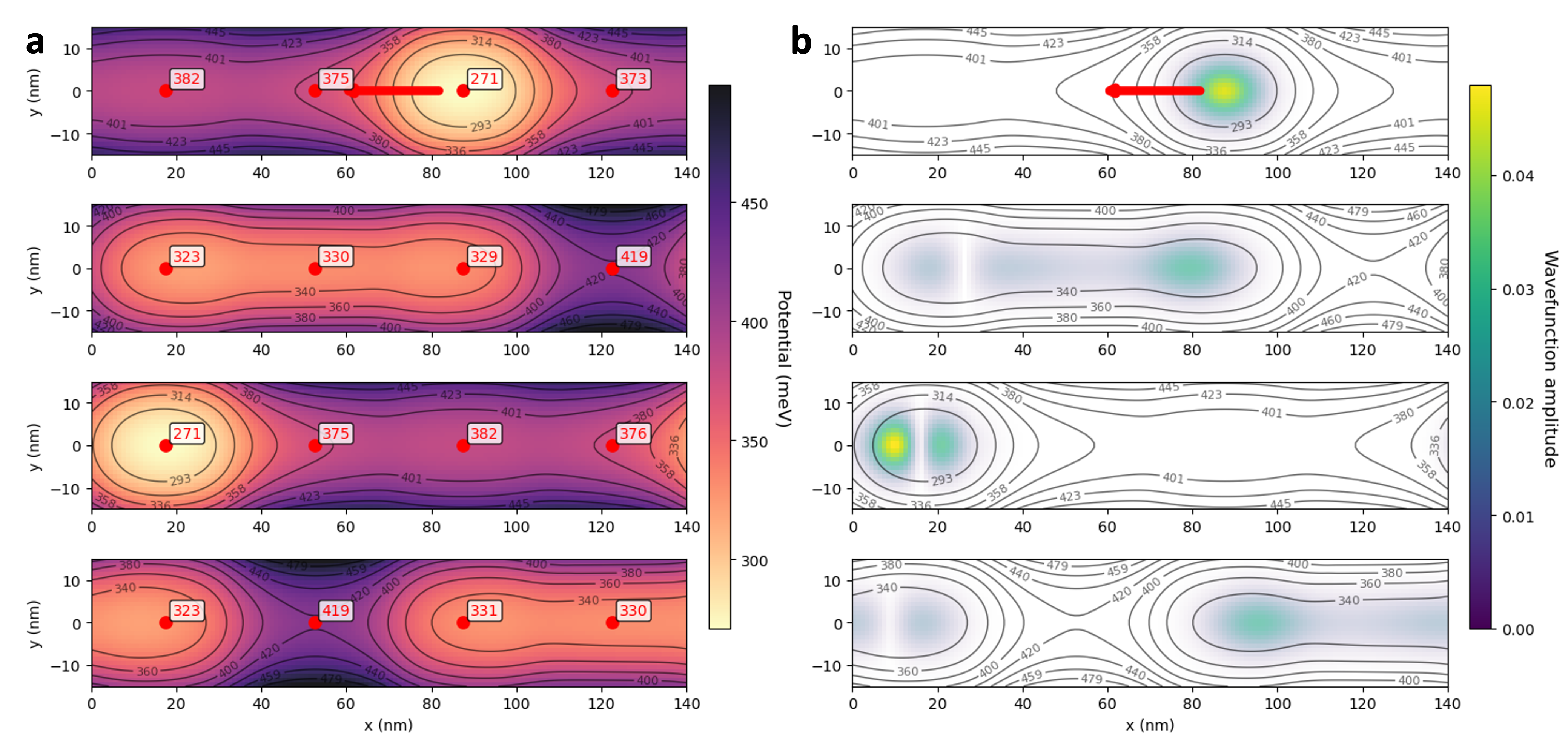}
    \caption{Snapshots from 3D simulations of shuttling 20~m/s along a flat interface with $V_c=100~\text{mV}$ and $V_s=-1500~\text{mV}$. Shown are slices in the x-y plane taken 1 nm below the oxide interface of (a) the potential generated by the gates and (b) the electron wavefunction with potential contours. The shuttling direction is indicated by the red arrows, and the potentials in the plane under clavier-gates ABCD with x-positions 17.5~nm, 52.5~nm, 87.5~nm, 122.5~nm are marked in red. Successive times at $t=0$, $t=\frac{1}{4}t_{\rm max}$, $t=\frac{1}{2}t_{\rm max}$, $t=\frac{3}{4}t_{\rm max}$ where $t_{\rm max}$ is the time for one complete shuttling period of 140~nm, are displayed top to bottom. At these voltages the confinement under layer 3 gates is very weak due to additional screening, and the conveyor minimum effectively jumps from under gate C to under gate A causing orbital excitations.}
    \label{fig:bucket_brigade_potential}
\end{figure*}

First, we perform simulations with a perfectly flat interface to study how different operating voltages and shuttling speeds impact shuttling for an idealised device. We vary both the shuttling speed ($v\in\{20,50,100,150,200,250\}~\text{m/s}$) and the clavier-gate AC voltage in Eq.~\eqref{eq: V_clavier} ($V_c\in\{100,150,200,250,500,1000\}~\text{mV}$), while keeping the side-gate voltage $V_s$ fixed at $-1500 ~\text{mV}$.

While we observe negligible charge loss of $P_L<10^{-2}$ across the parameter space, our simulations reveal a fundamental transition from the conveyor-belt shuttling mode to bucket-brigade shuttling around $V_{c}=100~\text{mV}$. As seen in Fig.~\ref{fig:low_confinement}a, at this voltage the final ground state fidelity approaches zero for all shuttling speeds. We identify the multilayer structure of the clavier-gates as the cause for this behaviour: gates B and D, deposited in layer 3 in our model, experience additional screening from 5~nm of oxide compared to the layer 2 gates (A and C). As seen in Fig.~\ref{fig:bucket_brigade_potential}, this produces substantially weaker confinement potentials under B and D ($\sim330~\text{meV}$) relative to gates A and C ($\sim270~\text{meV}$), effectively forming barriers in the channel. When the conveyor is below gates B or D in the shuttling cycle ($t=\frac{1}{4}t_{\rm max}$ and $t=\frac{3}{4}t_{\rm max}$ in Fig.~\ref{fig:bucket_brigade_potential}) the potential is insufficiently deep to trap the electron, causing it to tunnel across these gates. The resulting sudden change in the position of the potential minimum, shown in Fig.~\ref{fig:low_confinement}b, induces substantial orbital excitation of the electron.

Plotting the instantaneous eigenspectra levels (presented in Fig.~\ref{fig: levels} of Appendix~\ref{Appendix: levels}) reveals that when the B and D gates are the most positive, the energy levels of the confinement potential bunch closely together, enabling avoided level crossings and excitation. The sudden onset of the bucket-brigade regime can also be understood through the Landau-Zener formula \cite{Landau1932zurtheorie, Zener1932nonadiabatic, Stueckelberg1932theorie, Majorana1932atomi} for the diabatic transition probability, 
\begin{equation}\label{eq: landau-zener}
    P_D=e^{-2\pi\Gamma},
\end{equation}
where the exponent $\Gamma = a^2/(\hbar|\alpha|)$ is a function of the energy gap $a$ between levels and the rate of change of their separation $\alpha$. As the confinement weakens, the reduced energy gap dramatically increases $P_D$, making orbital excitations highly likely. Conversely, increasing the clavier voltage to $V_c\geq150~\text{mV}$ compensates for the additional screening of layer 3 gates sufficiently to suppress $P_D$ and restore conveyor-belt shuttling. The exponential nature of Landau-Zener transitions accounts for this sharp transition between the two operating regimes.

We also observe notable orbital excitation at $V_c=1000~\text{mV}$ where the ground-state fidelity drops to around $\sim90\%$, although this occurs for clavier voltages exceeding typical experimental operating voltages. We attribute this to the following mechanism: the stronger confinement resolves the discrete clavier-gate structure more acutely, leading to faster gate-to-gate transitions as the conveyor travels along the channel. These repeated velocity jumps increase $P_D$ by enough to cause partial excitations to higher orbital states in the lateral direction.

\subsection{Impact of Interface roughness}
\label{sec: results/interface roughness}

\begin{figure*}[t]
    \centering
    \includegraphics[width=1\textwidth]{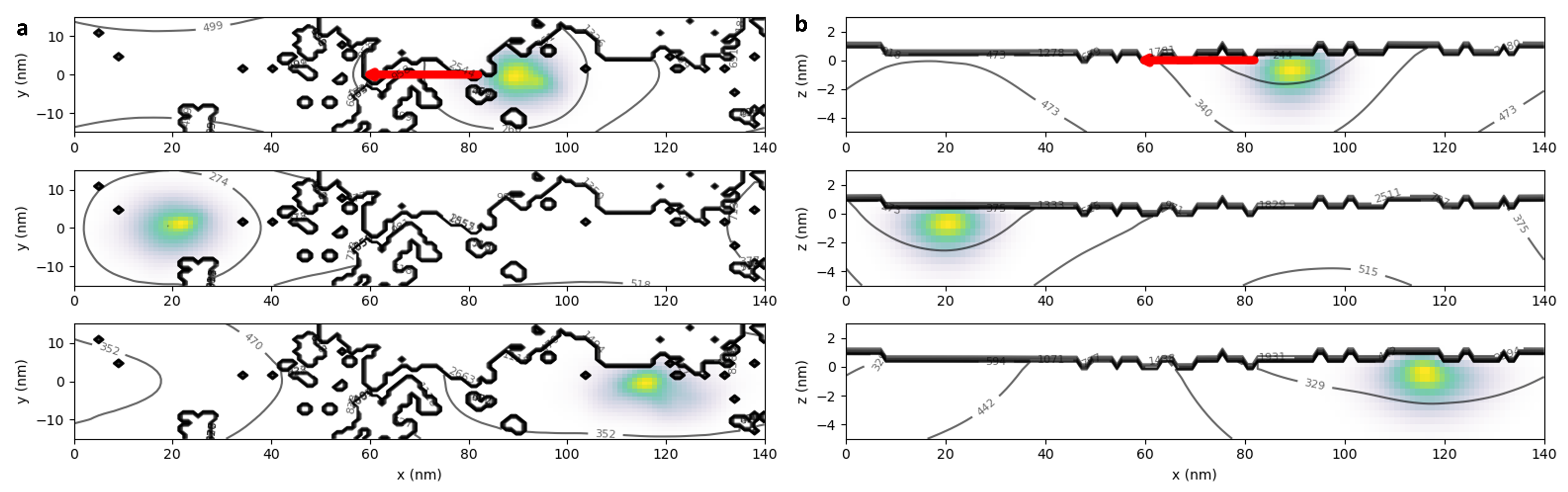}
    \caption{Snapshots of electron charge density from 3D simulations of shuttling an electron down the channel at 50~m/s with RMS=0.7~nm, $V_c=250~\text{mV}$ and $V_s=-1500~\text{mV}$. Shown are (a) a slice in the x-y plane taken 1~nm below the mean height of the oxide interface, and (b) a slice in the x-z plane taken at the center of the channel ($y=0\rm ~nm$). The shuttling direction is indicated by the red arrow, and contours of the potential (including the rough interface) in meV are shown in black. The thick black line corresponds to the 3000~meV potential step between the silicon bulk and silicon oxide layer. Successive times are shown, top to bottom. The model has periodic boundary conditions and so the electron exiting to the left in the middle frame, enters from the right in the lowest frame.}
    \label{fig:rms_frames}
\end{figure*}

Having identified the bucket-brigade regime in the ideal flat-interface case, we now examine how introducing interface roughness affects shuttling quality for different speeds and clavier-gate voltages. Successive snapshots from one of these simulations are shown in Fig.~\ref{fig:rms_frames}.

\begin{figure}[t!]
    \centering
    \includegraphics[width=1\linewidth]{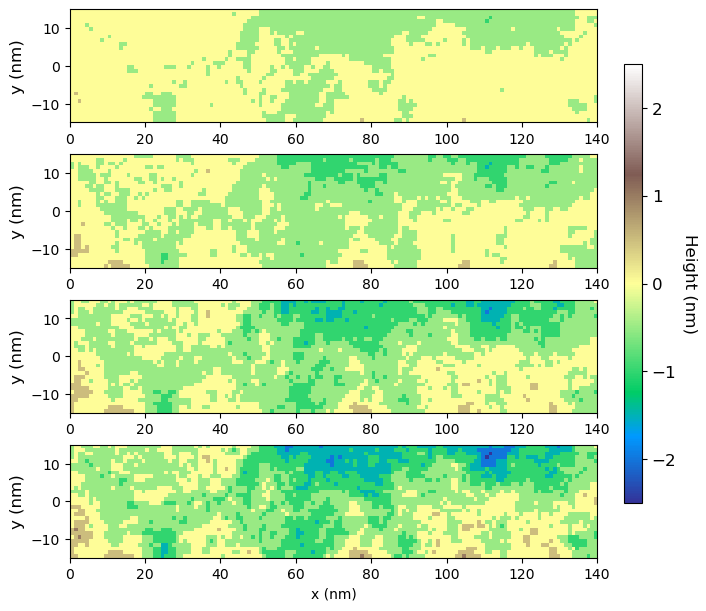}
    \caption{Interface topography of the simulated quantum region (marked in Fig.~\ref{fig: combined schematic}e) for the RMS values used in simulations. Interfaces for $\text{RMS}=0.3,0.5,0.7,0.9~\text{nm}$ using a Hurst coefficient of $H=0.3$ are displayed top to bottom. These were generated using the ACF-PSD approach discussed in Sec.~\ref{sec: interface_roughness}, where we model the interface using a power-law functional form for the ACF to capture the self-affine structure of the roughness.}
    \label{fig:surfaces}
\end{figure}

We use the ACF-PSD approach, discussed in Sec.~\ref{sec: interface_roughness}, to model the Si/SiO$_2$ interface for increasing root-mean-square roughnesses ($\text{RMS}\in\{0.3,0.5,0.7,0.9\}~\text{nm}$) shown in Fig.~\ref{fig:surfaces}. The Hurst coefficient $H$ of the roughness was fixed at 0.3 --- a value motivated by the interface topography study completed by Jacobs \textit{et al.} \cite{Jacobs2017surfacetopography}. 

To save computational resources, we perform our analysis using a single random interface realisation for a typical device rather than a statistical ensemble of different interfaces. This approach is justified by the self-averaging nature of the simulations: as the electron is shuttled over a distance of 140~nm, it samples numerous distinct topographical features.

\begin{figure*}[t!]
    \centering
    \includegraphics[width=1\textwidth]{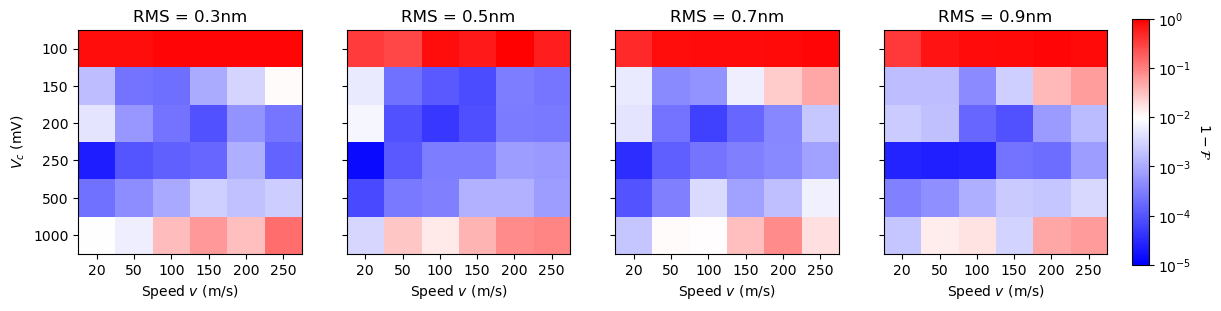}
    \caption{The orbital ground state infidelity at the end of the shuttling down the channel for different RMS values, clavier-gate voltages, and shuttling speeds, with the side-gate voltage fixed at $V_s=-1500~\text{mV}$. A diverging colourmap centered on a ground state fidelity of 99\% ($1-\mathcal{F}=10^{-2}$) indicates where shuttling becomes adiabatic. Considerable orbital excitation was observed at $V_c=100~\text{mV}$ across all roughnesses, demonstrating the onset of the bucket-brigade regime as discussed in Sec.~\ref{sec: results/low confinement}.}
    \label{fig:rms_heatmaps}
\end{figure*}

We observe negligible electron loss ($P_L<10^{-2}$) across the entire RMS-clavier voltage-shuttling speed parameter space simulated. Despite this, we find that the onset of the bucket-brigade regime is consistent at $V_c=100~\text{mV}$ for all roughnesses (see Fig.~\ref{fig:rms_heatmaps}), further highlighting that while standard electrostatic confinement suffices to localise an electron in SiMOS devices with realistic interface roughness, it may not be sufficient for the conveyor-belt shuttling mode for architectures where the clavier-gates are split across multiple gate layers.

Increasing the interface roughness does result in increased orbital excitations, but the effect remains modest: ground state fidelities stay above 99\% for most parameter combinations even with roughness as high as $\text{RMS}=0.9~\text{nm}$. For $\text{RMS}\geq0.7~\text{nm}$ and $V_c=150~\text{mV}$, shuttling speeds of 200~m/s or more were needed to partially excite the electron, reflecting the increased Landau-Zener transition probability when the system traverses avoided crossings more rapidly. Fortunately, typical shuttling speeds in Si/SiGe devices do not exceed $100~\text{m/s}$ \cite{desmet2025highfidelity}, and these RMS values are considerably higher even than those reported for interfaces from academic cleanrooms \cite{Cifuentes2024boundsto}, which are rougher than those fabricated by industrial foundries. Therefore, we conclude that roughness at the Si/SiO$_2$ interface does not present a significant challenge to charge shuttling.

\subsection{Impact of Gate Fabrication Imperfections}

The device fabrication process does not produce devices with perfect dimensions; scanning electron microscopy images of the gatestack in SiMOS devices reveal slight misalignments in gate centers and variations in gate widths \cite{lin2025interplayzeemansplittingtunnel}. To assess the robustness of the conveyor-belt shuttling scheme to this form of disorder, we simulate shuttling at 50~m/s with 10\%, 20\%, and 30\% variation in clavier-gate widths and gate-center positions across a range of voltages ($V_c\in\{250,500,750,1000\}~\text{mV}$).  Within the clavier-gate unit cell, this results in both a larger and a smaller gate. Therefore, if an electron can successfully transit both of these during one shuttling period, other realisations of misaligned gates within these bounds will likewise be tolerable.

\begin{figure*}[t!]
    \centering
\includegraphics[width=0.9\textwidth]{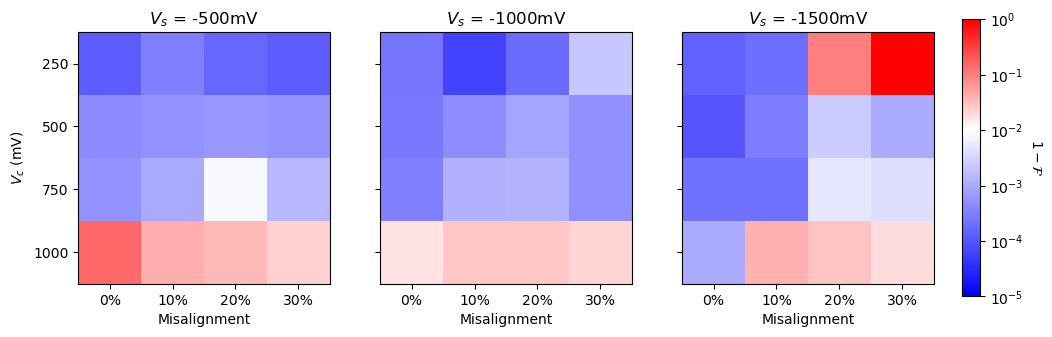}
    \caption{The orbital ground state infidelity at the end of shuttling in the x-direction at 100~m/s for different gate misalignments, clavier voltages and side-gate voltages. A diverging colourmap centered on a ground state fidelity of 99\% ($1-\mathcal{F}=10^{-2}$) indicates where shuttling becomes adiabatic. Excitation observed for $V_c=1000~\text{mV}$ is attributed to the electron more acutely resolving the misalignment in the discrete gate structure at stronger confinement.}
    \label{fig:misalignment_vs_C}
\end{figure*}

Across all cases tested, we observe negligible charge loss from the conveyor ($P_L<10^{-2}$) and ground state fidelities that are generally in excess of $\sim99\%$ (seen in Fig.~\ref{fig:misalignment_vs_C}), demonstrating that fabrication error tolerances up to 30\% gate misalignment are acceptable for shuttling devices without significantly compromising fidelity. This robustness can be understood through the smoothly varying potential landscape created by neighbouring clavier-gates, which averages out asymmetries introduced by gate misalignment and ensures the travelling potential minimum moves in a sufficiently continuous fashion.

However, increasing $V_c$ systematically reduces ground state fidelities to around $\sim90\%$ at $V_c=1000~\text{mV}$. Similarly, at very strong side-gate confinement ($V_s=-1500~\text{mV}$), misalignments of $20\%$ or greater cause the fidelity to suffer. This degradation occurs because stronger confinement produces smaller quantum dots that resolve the misaligned gate structure more acutely and undergo faster dot-to-dot transitions, increasing the diabatic transition probability $P_D$ in terms of the Landau-Zener framework.

Notably, at $V_s = -1500~\text{mV}$, fidelity exhibits non-monotonic dependence on $V_c$, with a minimum at $V_c = 250~\text{mV}$, improvement at $V_c = 500~\text{mV}$, and subsequent decrease at higher voltages. This behaviour reveals competing effects in the Landau-Zener transition probability given by Eq.~\eqref{eq: landau-zener}: weaker dot confinement reduces the orbital energy gap $a$ (increasing $P_D$) but simultaneously smooths the potential landscape to slow dot-to-dot transitions (decreasing $\alpha$ and so $P_D$).

Therefore, while optimal fidelity may require careful selection of confinement voltages, we can conclude that even in the case of substantial geometric irregularities of the clavier-gates, charge shuttling remains robust to this form of device imperfection.

\subsection{Impact of Charge defects}

Charge defects buried in the oxide or located at the interface are expected to present challenges for reliable electron shuttling in SiMOS devices \cite{Pillarisetty2021simosandsige}. In our simulations, we position single negative and positive charge defects on a plane perpendicular to the shuttling path, and investigate the impact of different operating voltages and shuttling speeds.

Here, we focus on the impact of quasi-static charge noise with dynamics that are slow relative to the shuttling operation. Our model represents monopole defects at the Si/SiO$_2$ interface ($P_b$ centers) or in the oxide ($E'$ centers) with energy levels sufficiently detuned from the gate and bulk Fermi levels to ensure that tunnelling rates remain negligible. While these defects may undergo small displacements via lattice deformations, the resulting fluctuations in the electrostatic environment are minimal. Faster charge fluctuators ($> \text{100 MHz}$) --- such as monopoles tunnel-coupled to the gate or bulk, or dipoles formed between coupled defects --- would introduce stochastic disturbances to our model that remain outside the scope of this study.

\subsubsection{Negative Charge Defects}

\begin{figure}[t]
    \centering
\includegraphics[width=1\linewidth]{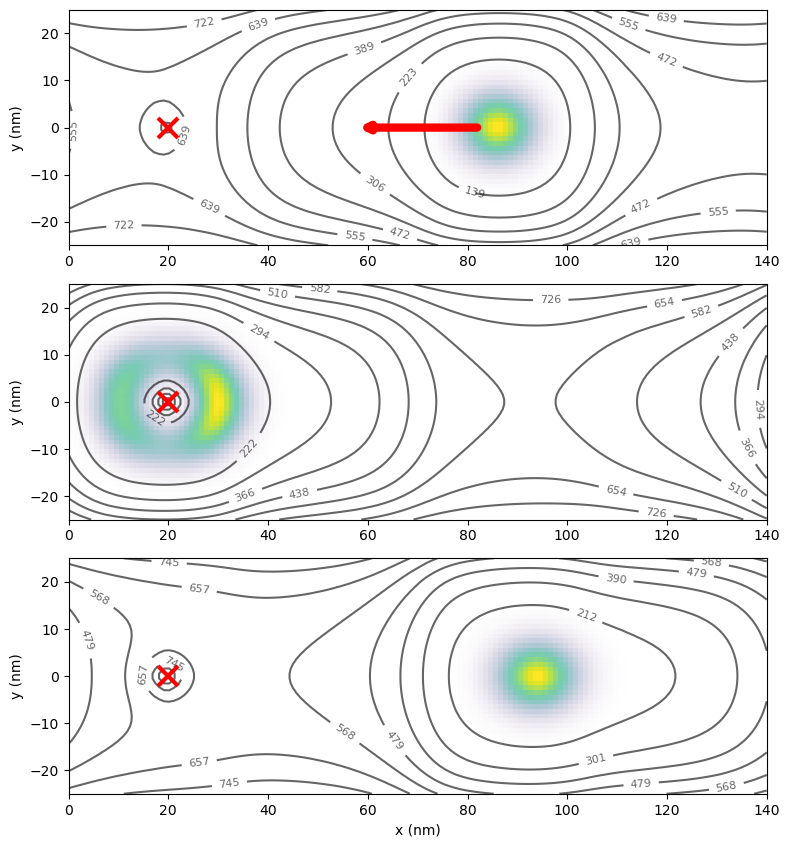}
    \caption{Snapshots from 3D simulations of shuttling an electron down a channel with a negatively charged defect located at the oxide interface (marked with a red cross). The shuttling speed was set to 100~m/s, and the clavier and side-gate voltages were set to $V_c=500~\text{mV}$ and $V_s=-1500~\text{mV}$. Shown is a slice taken 1~nm below oxide interface, with the shuttling direction indicated by the red arrow. Contour lines of the potential in meV are shown in black. Successive times are shown, top to bottom. The model has periodic boundary conditions and so the electron exiting to the left in the middle frame, enters from the right in the lowest frame. At these voltages, the defect leads to negligible orbital excitation and negligible charge loss.}
    \label{fig:neg_defect_frames}
\end{figure}

\begin{figure}[b]
    \centering
    \includegraphics[width=0.4\textwidth]{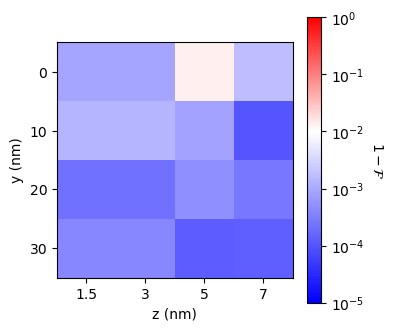}
    \caption{The orbital ground state infidelity at the end of the simulation for shuttling at 100 m/s along the x-direction with a negative charge defect at different positions in the channel. The clavier and side-gate voltages were fixed at $V_c=500~\text{mV}$ and $V_s=-1500~\text{mV}$. A diverging colourmap centered on a ground state fidelity of 99\% ($1-\mathcal{F}=10^{-2}$) indicates where shuttling becomes adiabatic.}
    \label{fig:neg_defect_x_vs_y}
\end{figure}

\begin{figure*}[t]
    \centering
    \includegraphics[width=0.8\textwidth]{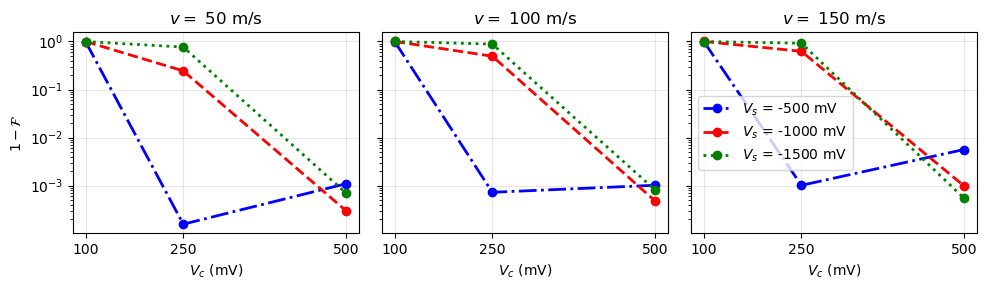}
    \caption{Orbital ground state infidelity at the end of shuttling with a negatively charged interface defect at the channel center, for varying clavier voltages ($V_c$), side-gate voltages ($V_s$), and shuttling speeds. Low $V_c$ induces bucket-brigade operation with near-complete transfer of population out of the ground state. At intermediate $V_c = 250$~mV, weak lateral confinement ($V_s = -500$~mV) allows the wavefunction to avoid the defect, suppressing excitation by 3-4 orders of magnitude. High $V_c$ maintains fidelity through large orbital energy gaps.}
    \label{fig:neg_defect_C_vs_B_vs_speed}
\end{figure*}

A negatively charged defect acts similarly to the barrier gate in a double quantum dot system, with the Coulomb repulsion forming a region of high potential in the channel. This effect is more pronounced for high side-gate voltages where the dot is more elongated.

For the position sweep, we place the negative defect at sixteen different points on the y-z plane perpendicular to the shuttling direction, and fix the shuttling speed to 100 m/s and clavier and side-gate voltages to $V_{c} = 500 ~\text{mV}$ and $V_s = -1500 ~\text{mV}$. Successive snapshots of one of these simulations are shown in Fig.~\ref{fig:neg_defect_frames}, and the ground state infidelity heatmaps for the entire parameter space are shown in Fig.~\ref{fig:neg_defect_x_vs_y}.

For these operating parameters and defect positions, we observe negligible charge loss from the conveyor ($P_L<10^{-2}$) even for the worst case of an interface defect located in the center of the channel. We also find that while the coulomb repulsion of the defect can deform the electron wavefunction substantially, shuttling remains almost entirely adiabatic for all defect positions with the ground state fidelity never dropping below $\sim99\%$.

However, a more nuanced picture emerges when we simulate shuttling while sweeping over different operating voltages. For these runs, we only consider an interface charge trap at the center of the channel. We shuttle the electron along the channel using nine combinations of clavier and side-gate voltages where $V_{c} \in \{100, 250, 500\} ~\text{mV}$ and $V_s \in \{-500, -1000, -1500\} ~\text{mV}$. The voltage sweep was repeated for three different shuttling speeds ($v= 50, 100, 150 ~\text{m/s}$). 

In most cases, lowering the clavier voltage below $500 ~\text{mV}$ causes a substantial decrease in the final ground state fidelity of the shuttled electron by several orders of magnitude. While we see ground state fidelities of $\sim99.9\%$ for high voltages, this number almost drops to zero for $V_{c}=100 ~\text{mV}$ (see Fig.~\ref{fig:neg_defect_C_vs_B_vs_speed}). An exception to this trend is observed for a side-gate voltage $V_s=-500 ~\text{mV}$, where the amount of orbital excitation initially decreased slightly at $V_c=250~\text{mV}$ and then increased significantly at $V_c=100~\text{mV}$ . Higher shuttling speeds have very little effect on the ground state fidelities, but we do observe some additional excitation caused by the increase in the Landau-Zener diabatic transition probability.

As discussed in Sec.~\ref{sec: results/interface roughness}, poor ground state fidelities at $V_c=100 ~\text{mV}$ are caused by unexpected bucket-brigade operation in the low voltage regime, although no doubt the presence of a negatively charge defect further perturbs the adiabatic transport pathway. The energy level spacings between instantaneous eigenstates are seen decreasing with $V_c$ in Fig.~\ref{fig: levels}b, making excitations to higher orbital states more likely according to Eq.~\eqref{eq: landau-zener}. For a larger clavier voltage of $V_c=250~\text{mV}$, while the confinement is sufficient for true conveyor-belt operation, it is not sufficient to suppress defect-induced orbital excitations resulting from these low energy splittings. The fact that reducing the side-gate voltage can reduce these excitations --- substantially, by three or four orders of magnitude to fidelities of $\sim99.9\%$--$99.99\%$, for $V_s=-500~\text{mV}$ --- is explained by the electron wavefunction being able to widen laterally in the proximity of the defect, minimising the overlap with the high energy region and so the perturbation strength. At $V_c=500~\text{mV}$, the higher confinement is such that even though the electron is forced into the proximity of the defect, the orbital energy gaps are large enough that only minor excitement occurs. Conversely, reducing the side-gate voltage here slightly increases the infidelity by reducing the energy spacing of the lateral modes, making excitation in these directions more likely.

\subsubsection{Positive Charge Defects}

\begin{figure}[t!]
    \centering
    \includegraphics[width=1\linewidth]{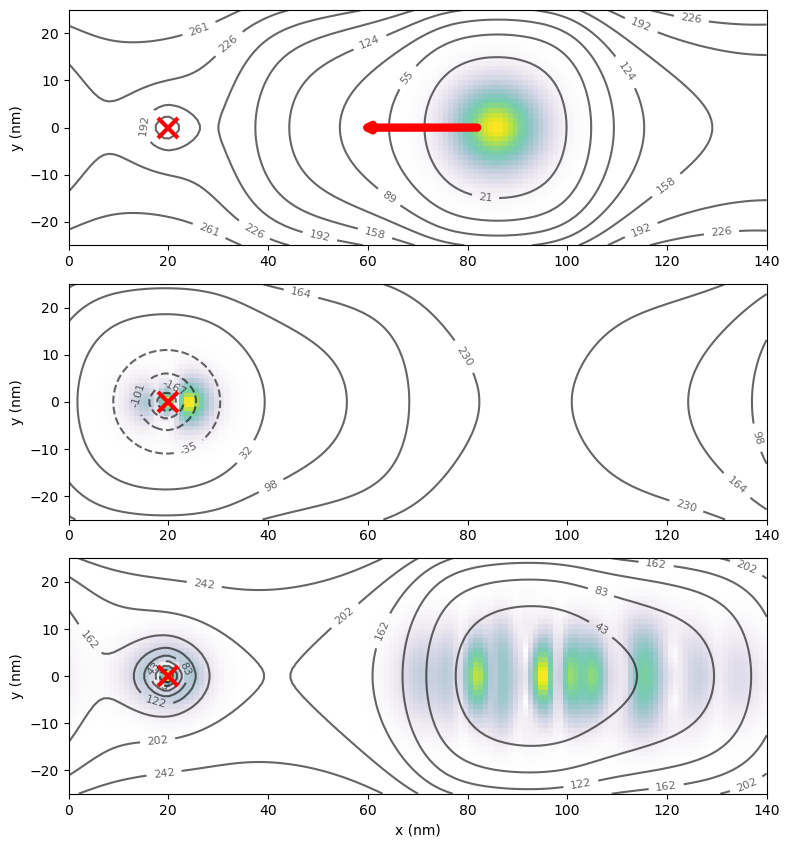}
    \caption{Snapshots from 3D simulations of shuttling an electron down a channel with a positively charged defect located at the interface (marked with a red cross). The shuttling speed was set to 50/s, and the clavier and side-gate voltages were set to $V_c=250~\text{mV}$ and $V_s=-500~\text{mV}$. Shown is a slice taken 1~nm below oxide interface, with the shuttling direction indicated by the red arrow. Contour lines of the potential in meV are shown in black. Successive times are shown, top to bottom. The model has periodic boundary conditions and so the electron exiting to the left in the middle frame, enters from the right in the lowest frame. At these voltages, the defect leads to considerable orbital excitation and partial charge loss, with 20\% of the electron wavefunction remaining trapped by the defect.}
    \label{fig:pos_defect_frames}
\end{figure}

\begin{figure}[b]
    \centering
    \includegraphics[width=0.4\textwidth]{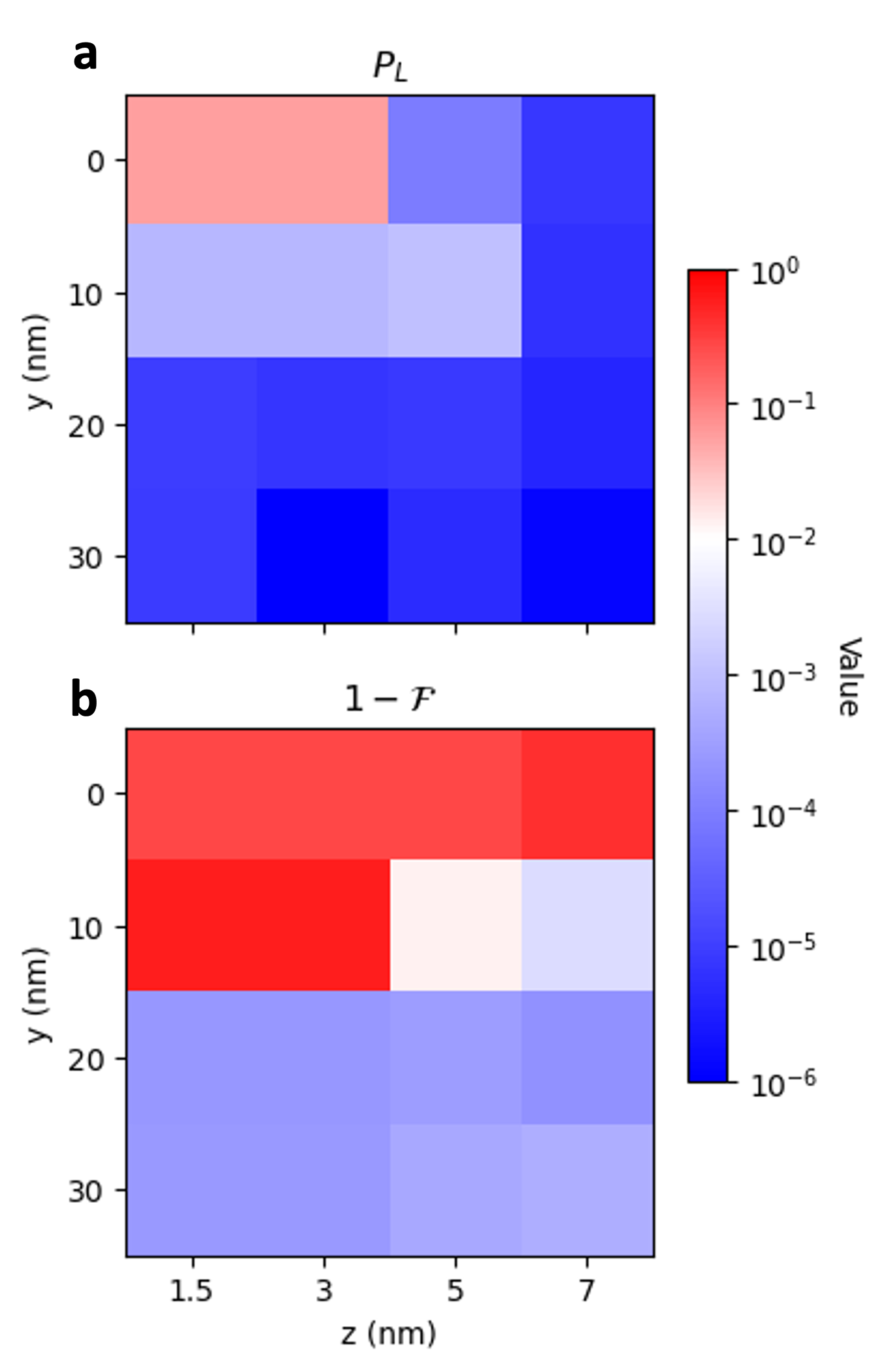}
    \caption{(a) The charge loss probability and (b) orbital ground state infidelity at the end of the simulation for shuttling at 100~m/s along the x-direction with a positive charge defect at different positions in the channel. The clavier and side-gate voltages were set to $V_c=500~\text{mV}$ and $V_s=-1500~\text{mV}$. A diverging colourmap centered on 99\% indicates where (a) electron loss becomes negligible or (b) shuttling becomes adiabatic.}
    \label{fig:pos_defect_y_vs_z}
\end{figure}

In contrast to negative defects, a single positively charged defect introduces an attractive potential for the electron that effectively creates an additional site for quantum dot formation. Our simulations show that the electron is permanently trapped at this site in extreme cases where the conveyor confinement is weak and the defect is located at the interface. Interestingly, the electron delocalises over the conveyor--trap system when the confinement is sufficient only for partial escape from the trap. Successive snapshots of one such case are shown in Fig.~\ref{fig:pos_defect_frames}. While lossless charge shuttling is restored by increasing the confinement, adiabaticity is only restored when the defect is moved from the center of the channel.

The position sweep, performed using $v=100~\text{m/s}$, $V_c=500$~mV and $V_s=-1500$~mV, demonstrates this sensitivity to defect position (see Fig.~\ref{fig:pos_defect_y_vs_z}). Partial loss of the electron wavefunction to the trap of $P_L\sim 5\%$ is observed for defects at $y=0 \rm~nm$, $z\leq3 \rm~nm$ only, and is negligible elsewhere ($P_L<10^{-2}$). Significant orbital excitations were consistently observed when the defect was located at the interface, or for a defect buried 3~nm into the oxide and  $\leq10~\text{nm}$ of the center of the channel. Defects positioned at 5 nm or more into the oxide were sufficiently screened so that they didn't lead to orbital excitations during shuttling.

\begin{figure*}[t!]
    \centering
    \includegraphics[width=0.8\textwidth]{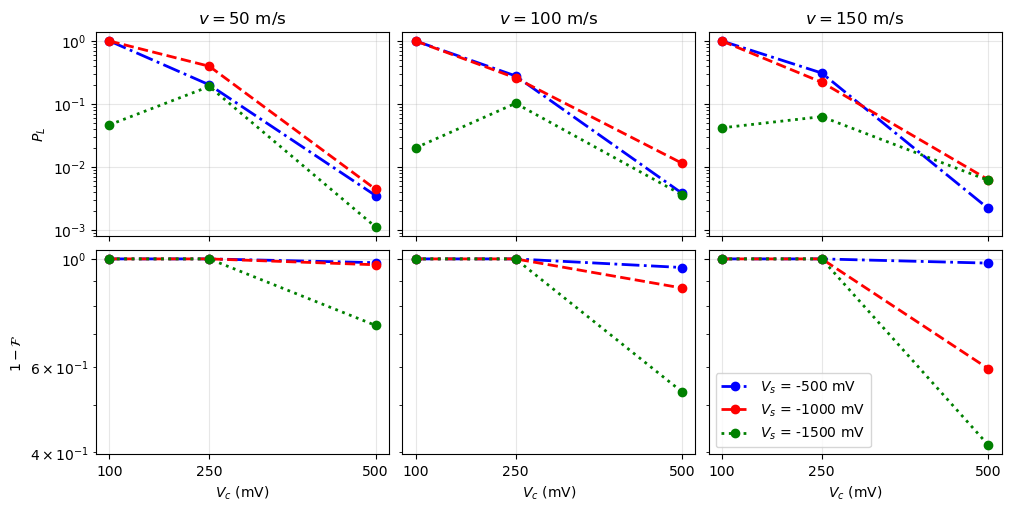}
    \caption{Charge loss probability (top panels) and orbital ground state infidelity (bottom panels) for shuttling with a positively charged defect at the channel center. At $V_c = 100$~mV, permanent electron capture occurs for $|V_s|\leq1000~\text{mV}$ and all shuttling speeds tested. The electron partially escapes the trap with an increased side-gate voltage of $V_s=-1500~\text{mV}$ or increased clavier-gate voltage of 250~mV, with $P_L$ values ranging from 5\%--30\%. At $V_c = 500$~mV, complete escape from the trap occurs but induces considerable orbital excitations ($\mathcal{F}=0\%-60\%$).}
    \label{fig:combined_probs_and_fids_pos_defect}
\end{figure*}

Meanwhile, the dependence on operating voltages reveals clear thresholds beyond which the conveyor confinement is insufficient to prevent the electron loss to the positive trap. Fig.~\ref{fig:combined_probs_and_fids_pos_defect} shows that, at the lowest clavier voltage ($V_c = 100~\text{mV}$), the shuttled electron is permanently trapped across side-gate voltages of $|V_s|\leq1000~\text{mV}$ and all speeds tested. Partial electron loss is observed when increasing the side-gate voltage to $V_s=-1500~\text{mV}$ or the clavier-gate voltage to 250~mV, with $P_L$ values ranging from 5\%--30\%. At $V_c=500~\text{mV}$, the conveyor confinement is strong enough for the electron to completely escape the trap for all side-gate voltages and shuttling speeds, however it emerges with considerable orbital excitation ($\mathcal{F}=0\%-60\%$).

Permanent localisation under the defect can be explained by analysing the evolution of the instantaneous eigenspectra of the combined conveyor-trap system, shown in Fig.~\ref{fig: levels}. We see a high concentration of sequential avoided crossings occurring when the conveyor begins to interact with the defect, with multiple bound states of the defect dropping deep below the conveyor levels. This forces the electron into a series of non-adiabatic transitions governed by Eq.~\eqref{eq: landau-zener} at these initial crossings, transferring its population to the deeply bound defect states. Once captured, the energy of the trapped electron remains far below the conveyor's evolving levels for the remainder of the simulation, maintaining a substantial energy gap of $\sim200~\text{meV}$. This effectively prevents any possibility of tunnelling back to the travelling conveyor minimum. For comparison, increasing the confinement with a clavier voltage of 500~mV substantially reduces the size of this energy gap as the conveyor moves away from the defect. As a result, the electron encounters another sequence of avoided crossings with the conveyor levels, enabling it to escape the trap as a superposition of excited conveyor eigenstates. Partial trapping at intermediate confinement is caused by the gap between the two competing potential wells reducing enough for only the higher energy trap states to form avoided crossings with the conveyor states, meaning just the excited components of the wavefunction transition back to the shuttling minima.

Even when the electron completely escapes the trap, considerable orbital excitations present a concern as they may open up spin-flip pathways, in particular those mediated by phonon-mediated relaxation of spin-orbit hybridised states \cite{ciroth2025numericalsimulationcoherentspinshuttling}. However, the probability of a spin-flip event will scale with the degree of orbital-spin hybridisation, which we know from reported g-factor variations is approximately $10^{-3}$ in these systems \cite{Cifuentes2024boundsto}. Therefore, spin-conserving decay channels are expected to remain dominant under realistic operating conditions for charge shuttling.

\section{Conclusions}
\label{sec: conclusions}

We evaluated conveyor-belt electron shuttling in Si/SiO$_2$ channels with interface roughness, gate-fabrication imperfections, and isolated charged defects across relevant speeds and operating voltages. A broad regime of operating points supports lossless, near-adiabatic transport over a full 140~nm period.  For reference, these are delineated in Table \ref{table: regimes} of Appendix \ref{Appendix: summary}. 

The near-adiabaticity would suggest that the Dykhne formula \cite{dykhne1962adiabatic} could be applied to our problem. However, this is not straight-forward since the instantaneous eigenstates are not available as analytic  functions of time, but rather in numerical form. Instead, the Landau-Zener formula is related to the time-dependant, instantaneous eigenspectra to qualitatively understand how the diabatic transition probability changes with gate voltages and shuttling speed in the presence of device imperfections (see Appendix \ref{Appendix: levels}).

Interface roughness does not cause charge loss; with moderate confinement we observe essentially perfect retention. Adiabaticity is governed by the operating point: at very low clavier voltage ($V_{c}=100~\text{mV}$) the system collapses into bucket-brigade dynamics due to extra screening of layer 3 gates, concentrating the system's instantaneous eigenlevels thereby increasing the likelihood of excitations --- even on an ideal flat interface. Raising $V_{c}$ to $\gtrsim 150~\text{mV}$ reopens the orbital gaps and restores conveyor-belt behaviour. On the other hand, very strong confinement of $\sim1000~\text{mV}$ can increase excitation by making the dot resolve the clavier-gate array more acutely leading to velocity-induced excitation during faster dot-to-dot transitions.

Furthermore, geometric imperfections are well tolerated. Shuttling at 50~m/s, up to 30\% variations in gate widths and centers produce no charge loss and only modest excitation for a wide range of operating voltages. For larger confinement ($V_s=-1500~\text{mV}$) where the electron benefits less from the smoothing out of irregularities in the gate structure, we observe an increase in velocity-induced orbital excitation for $\geq20\%$ misalignment. Fortunately, state-of-the-art SiMOS fabrication processes routinely achieve nanometer lithographic precision \cite{Li2020aflexible300nm} --- well within the tolerance of adiabatic conveyor-belt charge shuttling identified here.

Among the disorder mechanisms considered in this work, positive charge defects pose the greatest obstacle to shuttling in SiMOS. They form a competing well: at $V_{c}=100~\text{mV}$ the electron is captured by an interface trap in the center of the channel. Increasing $V_{c}$ shrinks the trap--conveyor energy gap and allows complete escape, though inducing orbital excitation in the process. An intermediate regime at $V_{c}=250~\text{mV}$, where the energy gap is only small enough for excited trap states to transition back to the conveyor levels, results in delocalisation over the combined trap--conveyor system. In contrast, single negative charges deform the wavefunction but do not result in charge loss, and only weakly affect fidelities, which remain high for appropriate operating voltages --- typically 99--99.99\%. Stronger excitations are observed when the electron cannot sufficiently deform around the negative trap due to strong lateral confinement.

While defect-induced excitations may open spin-flip pathways, the small spin-orbit admixture ($\sim 10^{-3}$) suggests spin-conserving decay remains dominant. In addition, the density of these positively charged interface defects --- the worst case --- can be reduced by advanced fabrication processes. Specifically, Hydrogen Resist Lithography (HRL), which utilises a buried hydrogen termination layer followed by annealing in N$_2$+H$_2$ forming gas, has been shown to significantly reduce the density of donor-like traps compared with pristine samples \cite{czarnecki2025hydrogenpassivationeffects}.

The modelling framework presented in this study is well suited for extension to larger parts of the Hilbert space. The inclusion of the spin and valley degrees of freedom, along with Linblad treatments capturing phonon-mediated relaxation, will enable a comprehensive assessment of quantum information transfer fidelity during electron shuttling in the presence of realistic device disorder. Interface roughness, for example, has been shown to strongly affect spin and valley physics \cite{Cifuentes2024boundsto, Losert2024strategiesforenhancingspin-shuttling, prentki2025boundsatomisticdisorderscalable}. Orbit- and valley-dependent g-factors lead to spin dephasing during shuttling, while population of the excited valley state after shuttling --- which constitutes leakage out of the computational subspace --- poses challenges for spin readout via Pauli spin blockade.

The inclusion of spin-orbit coupling in this framework will also enable its application to hole platforms, such as those confined in silicon and germanium heterostructures. Holes are characterised by large effective masses and strong spin-orbit coupling \cite{bosco2024holeshuttling}, which results in smaller orbital energy spacings. Given that the diabatic transition probability depends exponentially on the square of the energy gap, this is expected to enhance orbital excitations during shuttling significantly. This will likely shift the onset of the bucket-brigade regime to higher voltages as well as increase the susceptibility to charge defects. Notably, the effect of these defects is inverted for holes compared to electrons, and negative charge defects will present the biggest challenge by capturing passing holes. Additionally, the larger g-factor anisotropy may lead to greater dephasing whilst also suppressing spin-conserving relaxation channels. Detailed investigations using this modelling framework can thus help identify the optimal operating regimes for high-fidelity shuttling across a diverse range of spin-qubit platforms.

\section{Acknowledgements}

This research was supported by Innovate UK (IUK) under Grant No. 10149518 (SiQEC–Silicon Quantum Error Correction), funded through the Quantum Missions pilot: Quantum Computing and Quantum Networks competition.

\section{Data Availability}

The data that support the findings of this article are not publicly available because they are owned by a third party and the terms of use prevent public distribution. The data are available from the authors upon reasonable request.

\onecolumngrid % Switches to one-column mode

\newpage
\appendix

\section{Summary of Operational Regimes}\label{Appendix: summary}

Our simulations indicate that the clavier-gate voltage amplitude, $V_c$, is the primary control knob influencing shuttling performance. Based on this parameter, we identify three distinct operational regimes for charge shuttling within the model SiMOS device, outlined below. High fidelity (negligible loss $P_L < 1\%$ and adiabatic transit $1-\mathcal{F} < 1\%$) is achieved primarily in the high-voltage conveyor-belt (CV) regime:

\begin{itemize}
    \item \textbf{High $V_c$ ($\sim$ 500~mV):} Operates in the conveyor-belt mode where the Landau-Zener diabatic transition probability has been suppressed significantly, enabling adiabatic shuttling in most scenarios considered in our study. While they do not capture passing electrons, positive defects near the shuttling path still present a challenge by inducing orbital excitations, although we expect these excitations to relax rapidly via spin-conserving decay channels.
    \item \textbf{Moderate $V_c$ ($\sim$ 250~mV):} Sufficiently opens energy level spacings to restore conveyor-belt shuttling. Given suitable choices of shuttling speed and side-gate confinement ($V_s$), shuttling at this operating point is generally adiabatic in the presence of different device imperfections. For instance, negative defects deform passing electrons without impacting fidelities provided a weaker lateral confinement is used. However, positive charge defects positioned near the center of the channel at the interface still cause considerable excitations (orbital ground state fidelity $\mathcal{F}\ll90\%$), and can even partially capture the electron resulting in delocalisation over the combined defect-conveyor system.
    \item \textbf{Low $V_c$ ($\sim$ 100~mV):} Operates in an unsafe regime --- screening of the layer 3 gates results in unexpected bucket-brigade operation accompanied by significant orbital excitations. In this regime, the competing well formed by a positive charge defect in the channel can even permanently capture passing electrons.
\end{itemize}

\begin{table*}[h]

\centering
\caption{Summary of charge loss and orbital excitation regimes. These guidelines apply to shuttling speeds $\le 150$~m/s, with interface roughness up to $\text{RMS} = 0.9$~nm and gate misalignments up to 30\%. Operational bounds are determined by the worst-case defect placement within the center of the shuttling path at the interface. Shuttling infidelities with the orbital ground state $1-\mathcal{F} < 1\%$ are categorised as adiabatic, while probabilities of charge loss $P_L < 1\%$ (applicable with positive defects only) are categorised as ``lossless''.}
\begin{tabular}{@{}llllll@{}}
\toprule
\textbf{Clavier Voltage} & \textbf{Mode} & \textbf{Roughness} & \textbf{Misalignment} & \textbf{Negative Defects} & \textbf{Positive Defects} \\ \midrule
500~mV & CV & Adiabatic & Adiabatic & Adiabatic & Diabatic (Lossless) \\ 
250~mV & CV & Adiabatic & Adiabatic & Diabatic / Adiabatic\footnote{Adiabaticity is maintained at $V_c = 250$~mV provided a suitable choice of side-gate voltage ($V_s=-500$~mV); weaker lateral confinement allows the electron to deform around the defect potential, thereby avoiding orbital excitations.} & Diabatic (Partial Loss\footnote{Partial loss to the positive defect potential results in delocalisation over the combined defect-conveyor system.}) \\
100~mV & BB & Diabatic & Diabatic & Diabatic & Diabatic (Complete Loss) \\
\bottomrule
\label{table: regimes}
\end{tabular}
\end{table*}

\newpage
\section{Eigenenergy Level Diagrams}\label{Appendix: levels}

\begin{figure*}[h]
    \centering
    \includegraphics[width=1\textwidth]{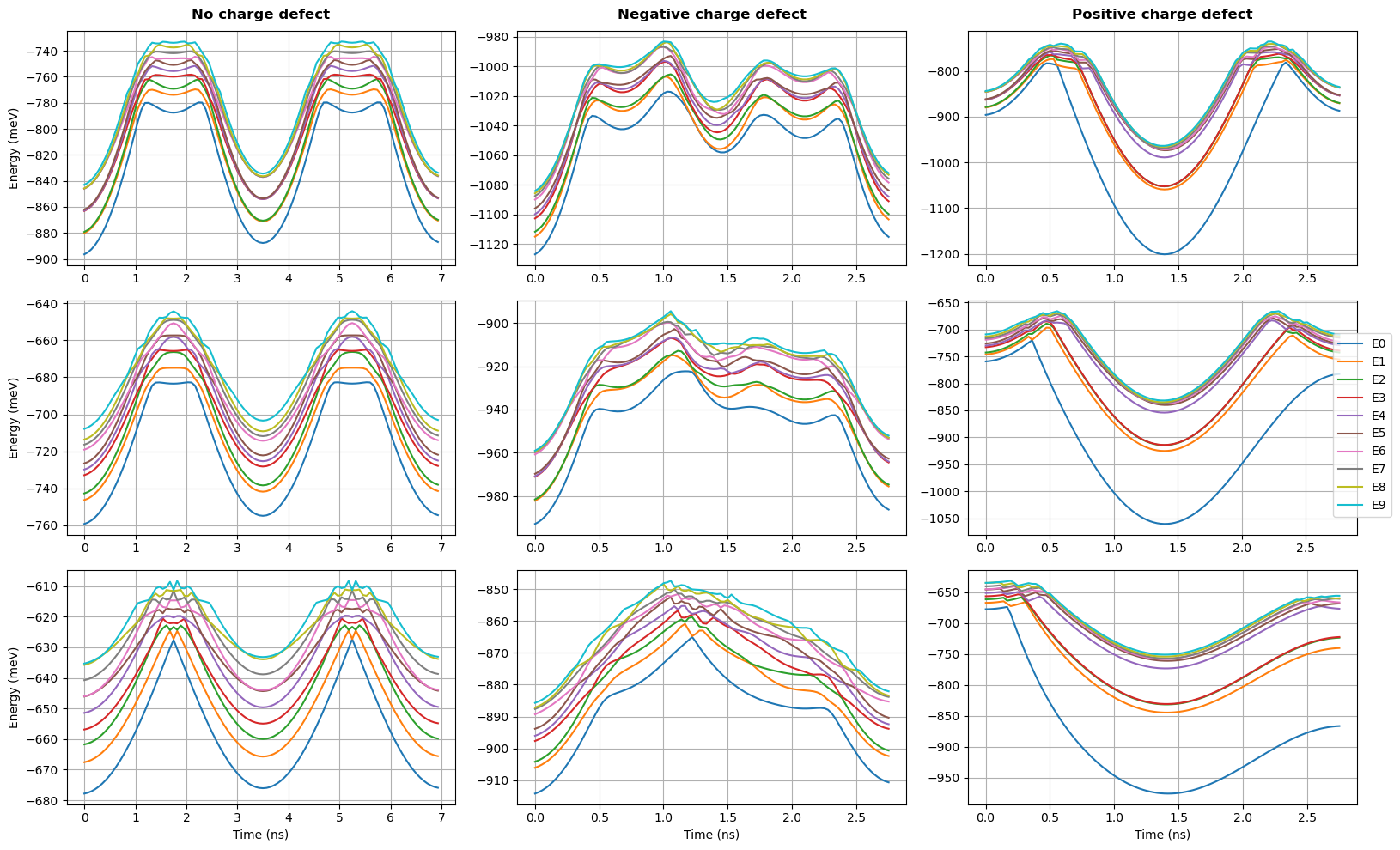}
    \caption{The 10 lowest instantaneous eigenenergies of the system potential plotted as a function of time for a different classes of shuttling simulation. The clavier voltage was set to (top) $V_c=500~\text{mV}$, (middle) $V_c=250~\text{mV}$, (bottom) $V_c=100~\text{mV}$. Stronger confinement opens up the energy gaps, making diabatic transitions governed by the Landau-Zener  (Eq.~\eqref{eq: landau-zener}) less likely. On the other hand, weaker confinement reduces the energy gaps leading to the formation of multiple avoided-crosses, greatly increasing the probability of diabatic transitions.}
    \label{fig: levels}
\end{figure*}

Analysing the instantaneous eigenspectra during a shuttling cycle can help us to understand the charge dynamics in the presence of different kinds of device disorder. In Fig.~\ref{fig: levels}, these are plotted for $V_c\in(500,250,100)~\text{mV}$ for different charge shuttling scenarios. Excitation is determined by the Landau-Zener formula for the diabatic transition probability:
\begin{equation}
    P_D=e^{-2\pi\Gamma},
\end{equation}
where the exponent $\Gamma = a^2/(\hbar|\alpha|)$ is a function of the energy gap $a$ between energy levels and the rate of change of their separation $\alpha$. For any given pair of levels, as the size of their energy gap $a$ is decreased by weakening the dot confinement or through the introduction of charge defects into the system, $P_D$ increases significantly. Similarly, shutting at higher speeds (higher sweep rate $\alpha$) increases the likelihood of orbital excitations.

\section{Solving the Poisson Equation}\label{Appendix: poisson}

To efficiently compute electrostatic potentials in our device, we exploit the linearity of the Poisson equation,
\begin{equation}
\nabla \cdot \left[ \epsilon(\mathbf{r}) \nabla \Phi(\mathbf{r}) \right] = -\rho(\mathbf{r}),
\end{equation}
where $\epsilon(\mathbf{r})$ is the spatially varying permittivity and $\rho(\mathbf{r})$ is the charge density. Because the equation is linear in both the boundary conditions and the source terms, the total potential landscape can be assembled from a set of precomputed Green’s functions.

We solve the Poisson equation using periodic boundary conditions in the shuttling (x) direction and Neumann boundary conditions in the y and z directions. This is done once for each gate electrode using a unit voltage applied to that gate while all others are held at zero. The resulting solution defines the gate’s Green’s function $G_i(\mathbf{r})$. Similarly, for each relevant point charge we compute a Green’s function $G_{\rho}(\mathbf{r}, \mathbf{r}_j)$, which captures the potential produced by a unit charge located at $\mathbf{r}_j$ with all gates grounded. These functions can then be linearly superposed to construct the full potential:
\begin{equation}
V(\mathbf{r}, t) 
= \sum_i V_i(t)\, G_i(\mathbf{r})
+ \sum_j Q_j\, G_{\rho}(\mathbf{r}, \mathbf{r}_j),
\end{equation}
where $V_i(t)$ is the applied voltage on gate $i$ and $Q_j$ is the magnitude of charge $j$.

To compute each Green’s function, we discretise the Poisson equation on a finite grid and solve it iteratively using the Gauss--Seidel method with overrelaxation. At each iteration, the potential is updated according to
\begin{equation}
\Phi^{(k+1)} 
= (1-\omega)\, \Phi^{(k)} 
+ \omega\, \Phi^{\text{new}},
\end{equation}
where $\Phi^{\text{new}}$ is the locally updated potential and $\omega$ is the relaxation parameter chosen to optimize convergence.

This approach requires solving the Poisson equation only once per gate and once per charge position. Hence the subsequent evaluation of the potential landscape is computationally inexpensive.

\section{Solving the Schrödinger Equation}\label{Appendix: schrodinger}

The dynamics of a single electron are governed by the time-dependent Schr\"odinger equation
\begin{equation}
\label{schrodinger}
i\hbar \frac{\partial}{\partial t}\ket{\psi(t)} = H(t)\ket{\psi(t)},
\end{equation}
which we solve for the time-dependent potentials generated in Sec.~\ref{sec: potential landscape}. In this work, we implemented two complementary numerical approaches for solving Eq.~\eqref{schrodinger}: a direct split-operator method used as a high-accuracy benchmark, and a more efficient spectral projection method used for large-scale shuttling simulations.

\begin{figure*}[t]
    \centering
    \includegraphics[width=1\textwidth]{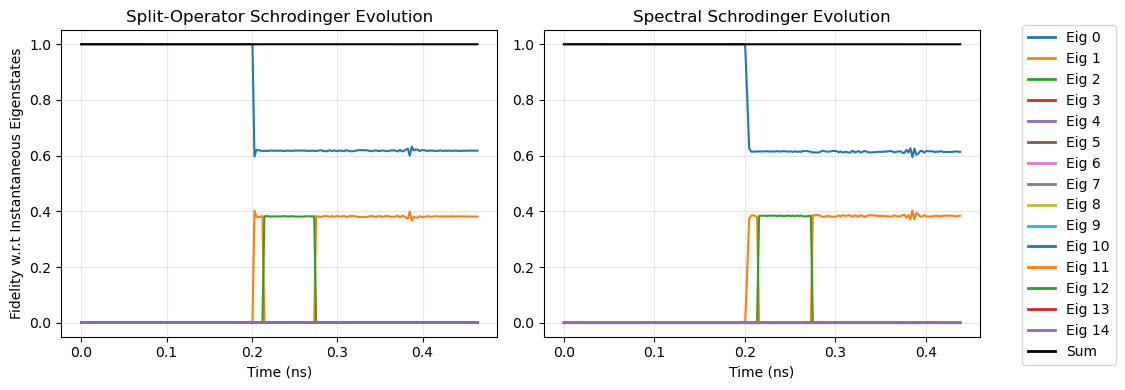}
    \caption{Evolution of squared level coefficients $|c_m(t)|^2$ using the split-operator and spectral projection methods for performing Schrödinger evolution, demonstrating excellent agreement between the two approaches for studying orbital excitations. Here, an electron was shuttled at 300 m/s down a channel with a negatively charged defect in the oxide with $V_c=250~\text{mV}$ and $V_s=-1000\text{mV}$. For this scenario the split-operator method took $\sim3$ days to compute while the spectral method took $\sim1$ hour --- a speedup of $65\times$.}
    \label{fig: neg defect validation}
\end{figure*}

To validate the accuracy of the faster projection method, we compared the level coefficients obtained using both numerical approaches for adversarial shuttling trajectories that induce significant orbital excitation. The two approaches exhibit excellent quantitative agreement, illustrated in Fig.~\ref{fig: neg defect validation}, confirming that the reduced subspace captures the relevant physics with negligible loss of accuracy.

\subsection{Split-Operator Time Evolution}
\label{Appendix:split_operator_time_evolution}

The split-operator method approximates the short-time propagator via Trotter factorisation,
\begin{align}
\ket{\psi(t+dt)}
&= e^{-iH(t) dt/\hbar}\ket{\psi(t)} \\
&\approx e^{-iV dt/2\hbar}\, e^{-iT dt/\hbar}\, e^{-iV dt/2\hbar}\ket{\psi(t)}
+ \mathcal{O}(dt^{3}),
\label{eqs: trotter}
\end{align}
where $H = T + V$ and Strang splitting yields a second-order accurate approximation \cite{strang1968}.  
To resolve the rapidly oscillating dynamical phase, simulations require femtosecond-scale timesteps. While this method is highly accurate, it is computationally expensive and impractical for parameter sweeps, and we therefore use it primarily as a benchmark.

\subsection{Spectral Projection Method}
\label{Appendix:projection-method}

To efficiently simulate realistic shuttling trajectories, we use a spectral projection scheme that separates two dynamical timescales: the rapid intrinsic phase evolution and the slow evolution of the instantaneous eigenbasis as the potential changes. At each timestep, we compute the instantaneous eigenstates $\ket{\phi_n(t)}$ and eigenenergies $\epsilon_n(t)$ of $H(t)$ using a Lanczos eigensolver \cite{arpack}.

The Hamiltonian is applied in a matrix-free form via FFTs (denoted by $\mathcal{F}$):
\begin{equation}
\label{fourierHamiltonian}
H(t) = \mathcal{F}^{-1} T \mathcal{F} + V(t),
\end{equation}
where the kinetic operator,
\begin{equation}
T =
\frac{\hbar^2}{2}
\left(
\frac{k_x^2}{m_t}
+ \frac{k_y^2}{m_t}
+ \frac{k_z^2}{m_\ell}
\right),
\end{equation}
is diagonal in momentum space, enabling efficient application to real-space wavefunctions.

Writing $\ket{\psi(t)} = \sum_m c_m(t)\ket{\phi_m(t)}$, the evolution is given by
\begin{align}
c_n(t+dt)
=
\sum_m
O_{nm}\,
e^{-i\epsilon_m(t) dt/\hbar}\,
c_m(t),
\end{align}
where
\begin{equation}
O_{nm} = \braket{\phi_n(t+dt)|\phi_m(t)}
\end{equation}
is the overlap matrix between successive eigenbases. A derivation and proof of convergence to exact Schr\"odinger evolution for $dt \to 0$ are provided in Appendix~\ref{Appendix: subspace derivation}.

\section{Derivation of the Spectral Projection Method}\label{Appendix: subspace derivation}

\subsection{Exact Schrödinger Evolution}

The exact single-particle time-evolution is governed by the time-dependent Schrödinger equation:
\begin{equation}
i\hbar \frac{\partial}{\partial t}\ket{\psi(t)} = H(t)\ket{\psi(t)},
\end{equation}
where the formal solution is given by the time-evolution operator:
\begin{equation}
\ket{\psi(t)} = U(t,t_0)\ket{\psi(t_0)}.
\end{equation}
We can find the instantaneous eigenstates at each instant $t$ by diagonalising the Hamiltonian:
\begin{equation}
H(t)\ket{\phi_n(t)} = \epsilon_n(t)\ket{\phi_n(t)}.\label{eq: eigenvalue}
\end{equation}
These states form a complete orthonormal basis:
\begin{equation}
\sum_n \ket{\phi_n(t)}\bra{\phi_n(t)} = \mathbb{I}, \quad \braket{\phi_n(t)|\phi_m(t)} = \delta_{nm},
\label{eq:ONB}
\end{equation}
and expanding the wavefunction in this basis gives
\begin{equation}
\ket{\psi(t)} = \sum_n c_n(t)\ket{\phi_n(t)},\label{eq: expansion}
\end{equation}
where $c_n(t) = \braket{\phi_n(t)|\psi(t)}$ are time-dependent, complex coefficients. To get an equation for the evolution of the coefficients in the instantaneous eigenbasis, we substitute the expansion \eqref{eq: expansion} into the Schrödinger equation:
\begin{equation}
i\hbar \frac{\partial}{\partial t}\left(\sum_n c_n(t)\ket{\phi_n(t)}\right) = H(t)\sum_n c_n(t)\ket{\phi_n(t)}.
\end{equation}
Applying the product rule on the left-hand side, and substituting the eigenvalue equation \eqref{eq: eigenvalue} on the right-hand side of the equation leads to:
\begin{equation}
i\hbar \sum_n \left[\dot{c}_n(t)\ket{\phi_n(t)} + c_n(t)\ket{\dot{\phi}_n(t)}\right] = \sum_n c_n(t)\epsilon_n(t)\ket{\phi_n(t)}.
\end{equation}
Projecting both sides of the equation onto $\bra{\phi_m(t)}$ and using \eqref{eq:ONB} gives
\begin{equation}
i\hbar \dot{c}_m(t) + i\hbar \sum_n c_n(t)\braket{\phi_m(t)|\dot{\phi}_n(t)} = c_m(t)\epsilon_m(t),
\end{equation}
and then rearranging for $\dot{c}_m(t) $ gives the exact equation of motion for the coefficients:
\begin{equation}
    \dot{c}_m(t) = -\frac{i}{\hbar}\epsilon_m(t)c_m(t) - \sum_n c_n(t)\braket{\phi_m(t)|\dot{\phi}_n(t)}.\label{eq: coefficient evolution}
\end{equation}
The first term is associated with the dynamical phase, which governs the rapid, state-specific phase oscillation. The second term contains both the Berry connection and the non-adiabatic coupling, capturing the transitions between basis states that are caused by the eigenbasis rotating in Hilbert space. This can be rearranged further to distinguish these different contributions:
\begin{equation}
    \dot{c}_m(t) = \underbrace{-\frac{i}{\hbar}\epsilon_m(t)c_m(t)}_{\rightarrow\text{Dynamical Phase}} -\underbrace{\braket{\phi_m(t)|\dot{\phi}_m(t)}c_m(t)}_{\rightarrow\text{Berry Phase}} - \underbrace{\sum_{n\neq m} \braket{\phi_m(t)|\dot{\phi}_n(t)}c_n(t)}_{\rightarrow\text{Non-adiabatic Coupling}}.
\end{equation}

\subsection{Evolution with the Spectral Projection Method}

With the spectral projection method, we discretise time ($t \to t + dt$) and propagate the wavefunction by iteratively applying (exactly, via exponentiation) the above dynamical phase and projecting onto the next eigenbasis. The first step is given by
\begin{equation}
    \ket{\psi^{(\text{phase})}(t+dt)} = \sum_{m \in \mathcal{S}} e^{-i\epsilon_m(t)dt/\hbar}c_m(t)\ket{\phi_m(t)},
\end{equation}
where $\mathcal{S}$ is a truncated subset of instantaneous eigenstates rather than the complete basis, which significantly reduces the computational cost. For brevity, we will drop the $\mathcal{S}$ and assume we are always working with an eigenbasis truncated at the highest non-zero $c_m$ over the full evolution. Next, projecting $\ket{\psi^{(\text{phase})}(t+dt)}$ onto the new (truncated) eigenbasis:
\begin{equation}
    c_n(t+dt) = \braket{\phi_n(t+dt)|\psi^{(\text{phase})} (t+dt)} = \sum_{m} O_{nm} e^{-i\epsilon_m(t)dt/\hbar}c_m(t)
\end{equation}
where $O_{nm} = \braket{\phi_n(t+dt)|\phi_m(t)}$
is the overlap matrix between the old and new basis states. The state after one timestep is then

\begin{equation}
    \ket{\psi(t+dt)} = \sum_{n} c_n(t+dt)\ket{\phi_n(t+dt)} = \sum_{nm} e^{-i\epsilon_m(t) dt / \hbar} O_{nm}c_m(t)\ket{\phi_n(t+dt)},\label{eq: time-evolved}
\end{equation}
or, written in terms of the initial wavefunction by substituting $c_m(t)=\braket{\phi_m(t)|\psi(t)}$,

\begin{equation}
\ket{\psi(t+dt)} = \sum_{nm} O_{nm} e^{-i\epsilon_m(t) dt / \hbar} \ket{\phi_n(t+dt)}\braket{\phi_m(t)|\psi(t)}.\label{eq: spectral}
\end{equation}

\subsection{Equivalence to Exact Schr\"odinger Evolution}

In the limit of $dt \to 0$, the spectral projection method is equivalent to exact Schrödinger evolution. To show this, first we expand the dynamical phase factor:

\begin{equation}
    e^{-i\epsilon_m(t)dt/\hbar} = 1 - \frac{i}{\hbar}\epsilon_m(t)dt + \mathcal{O}(dt^2).
\end{equation}
Next, we relate the overlap matrix $O_{nm}$ in Eq.~\eqref{eq: spectral} to the non-adiabatic coupling term in the equation of motion of the coefficients \eqref{eq: coefficient evolution}. To do this, we rewrite $O_{nm}$ by also expanding $\ket{\phi_n(t+dt)}$ around $t$ using a Taylor series:

\begin{equation}
    \ket{\phi_n(t+dt)} = \ket{\phi_n(t)} + dt \ket{\dot{\phi}_n(t)} + \mathcal{O}(dt^2).
\end{equation}
Substituting this into $O_{nm} = \braket{\phi_n(t+dt)|\phi_m(t)}$:

\begin{align}
  O_{nm} &= \left(\bra{\phi_n(t)} + dt \bra{\dot{\phi}_n(t)} + \mathcal{O}(dt^2)\right)\ket{\phi_m(t)}  \\
  &= \underbrace{\braket{\phi_n(t)|\phi_m(t)}}_{\delta_{nm}} + dt \underbrace{\braket{\dot{\phi}_n(t)|\phi_m(t)}}_{-\braket{\phi_n(t)|\dot{\phi}_m(t)}} + \mathcal{O}(dt^2)
\end{align}
where we have used the orthogonality of the basis ($\delta_{nm}$) and the product rule for the time-derivative term, highlighted below:

\begin{align}
    \frac{d}{dt}\braket{\phi_n(t)|\phi_m(t)} &= 0 \\
    \implies \braket{\dot{\phi}_n(t)|\phi_m(t)} &= - \braket{\phi_n(t)|\dot{\phi}_m(t)}.
\end{align}
Combining, we arrive at an expression for the overlap matrix:
\begin{equation}
    O_{nm} = \delta_{nm} - dt \braket{\phi_n(t)|\dot{\phi}_m(t)} + \mathcal{O}(dt^2).
\end{equation}
The diagonal elements of this matrix represent the state staying in the same instantaneous eigenstate, and off-diagonal terms that are directly proportional to the non-adiabatic coupling term from the exact coefficient equation of motion \eqref{eq: coefficient evolution}, capturing transitions between instantaneous eigenstates.

Returning to Eq.~\eqref{eq: spectral} and substituting the approximate expressions for the phase factor and overlap matrix, we arrive at

\begin{equation}
    c_n(t+dt) \approx \sum_m \left(\delta_{nm} - dt \braket{\phi_n|\dot{\phi}_m}\right)\left(1 - \frac{i}{\hbar}\epsilon_m dt\right)c_m(t).
\end{equation}
Performing the multiplication inside the summation, dropping terms of $\mathcal{O}(dt^2)$ and higher (denoted by $\dots$ below), and simplifying using the Kronecker delta, gives:
\begin{align}
    c_n(t+dt) &\approx \sum_m \left[\delta_{nm} - \delta_{nm}\frac{i}{\hbar}\epsilon_m dt - dt \braket{\phi_n|\dot{\phi}_m} + \dots\right]c_m(t), \\
     &\approx c_n(t) + \left[ -\frac{i}{\hbar}\epsilon_n(t)c_n(t) - \sum_m c_m(t)\braket{\phi_n(t)|\dot{\phi}_m(t)} \right]dt.
\end{align}
Finally, we rearrange this to obtain the finite-difference approximation of the time derivative:

\begin{equation}
    \frac{c_n(t+dt) - c_n(t)}{dt} \approx -\frac{i}{\hbar}\epsilon_n(t)c_n(t) - \sum_m c_m(t)\braket{\phi_n(t)|\dot{\phi}_m(t)}
\end{equation}
In the limit $dt \to 0$, the LHS becomes $\dot{c}_n(t)$, and the result is:

\begin{equation}
    \dot{c}_n(t) \approx -\frac{i}{\hbar}\epsilon_n(t)c_n(t) - \sum_m c_m(t)\braket{\phi_n(t)|\dot{\phi}_m(t)}
\end{equation}
This confirms that the spectral projection method is equivalent to the exact evolution equation \eqref{eq: coefficient evolution} in the limit ($dt \to 0$).

\subsection{Applicability of the Spectral Projection Method}
\label{Appendix:projection-applicability}

The projection method reproduces exact Schr\"odinger evolution in the limit $dt \rightarrow 0$.  
In practice, however, we use timesteps on the order of $\sim 1\,\mathrm{ps}$, which is many orders of magnitude larger than the femtosecond-scale steps required by a split-operator scheme.  
Despite this, the method remains accurate in the regimes relevant for electron shuttling, which can be understood by examining the evolution of the instantaneous eigenstate coefficients,
\begin{equation}
    \dot{c}_m(t) \approx \underbrace{-\frac{i}{\hbar}\epsilon_m(t)c_m(t)}_{\rightarrow\text{Dynamical Phase}} - \underbrace{\sum_{n} c_n(t)\braket{\phi_m(t)|\dot{\phi}_n(t)}}_{\rightarrow\text{Non-adiabatic Coupling}}.
\end{equation}

\paragraph*{Slowly varying Hamiltonian.}
When $H(t)$ evolves slowly compared with the instantaneous level spacings, the non-adiabatic couplings  
$\braket{\phi_m(t)|\dot{\phi}_n(t)}$ are small.  
In this regime, the projection method is highly accurate because the eigenbasis rotates smoothly and the coefficients $c_m(t)$ evolve predominantly through the (exactly captured) dynamical phase factor $e^{-i\epsilon_m dt/\hbar}$.  
Thus, the method successfully describes adiabatic motion of the electron along the shuttling trajectory.

\paragraph*{Rapid Hamiltonian changes and level crossings.}
At the opposite extreme, when the potential changes rapidly --- such as during avoided or near-avoided crossings --- the instantaneous eigenbasis can rotate significantly between timesteps.  
Here the projection method remains accurate because the overlap matrix $O_{nm}$ captures exactly how population is redistributed between instantaneous eigenstates.  
In the limit of an abrupt change, where the Hamiltonian jumps faster than the system can respond, the method automatically performs the correct sudden-basis projection: the state immediately after the jump is simply the projection of the preceding wavefunction onto the new eigenbasis.  
Thus the projection method naturally reproduces sudden-limit physics.

\twocolumngrid % Switches to two-column mode

% \bibliography{references} % For submission, replace this line with the output.bbl text


\end{document}